\documentclass{aa}
\usepackage{graphicx}
\usepackage{txfonts}
\usepackage{booktabs}
\usepackage{siunitx}
\usepackage{hyperref}
\usepackage{longtable}
\usepackage{amsmath}
\usepackage{upgreek}
\usepackage{multicol}

\begin{document}
\title{VLBI observations of GRB~201015A, a relatively faint GRB with a hint of Very High Energy gamma-ray emission}
\subtitle{}
\author{S.~Giarratana \inst{1,2}\fnmsep\thanks{email:stefano.giarratana2@unibo.it} \and L.~Rhodes \inst{3,4} \and B.~Marcote \inst{5} \and R.~Fender \inst{3,6} \and G.~Ghirlanda \inst{7,8} \and M.~Giroletti \inst{2} \and L.~Nava \inst{7,9,10} \and J.~M.~Paredes \inst{11} \and M.~E. Ravasio \inst{7,12,17} \and M.~Rib\'o \inst{11,13} \and M.~Patel \inst{14} \and J.~Rastinejad \inst{15} \and G.~Schroeder \inst{15} \and W.~Fong \inst{15} \and B.~P.~Gompertz \inst{16} \and A.J.~Levan \inst{17,18} \and P.~O'Brien \inst{14}}

\institute{Department of Physics and Astronomy, University of Bologna, via Gobetti 93/2, 40129 Bologna, Italy
\and
INAF Istituto di Radioastronomia, via Gobetti 101, 40129 Bologna, Italy
\and
Astrophysics, Department of Physics, University of Oxford, Keble Road, Oxford OX1 3RH, UK
\and
Max-Planck-Institut f{\"u}r Radioastronomie, Auf dem H{\"u}gel 69, 53121 Bonn, Germany
\and
Joint Institute for VLBI ERIC, Oude Hoogeveensedijk 4, 7991~PD Dwingeloo, The Netherlands
\and
Department of Astronomy, University of Cape Town, Private Bag X3, Rondebosch 7701, South Africa
\and
INAF Osservatorio Astronomico di Brera, Via E. Bianchi 46, I-23807 Merate, Italy
\and
INFN Sezione di Milano-Bicocca, Piazza della Scienza 3, I-20126 Milano, Italy
\and
INFN Sezione di Trieste, via Valerio 2, I-34149 Trieste, Italy
\and
Institute for Fundamental Physics of the Universe (IFPU), I-34151 Trieste, Italy
\and
Departament de F\'isica Qu\`antica i Astrof\'isica, Institut de Ci\`encies del Cosmos, Universitat de Barcelona, IEEC-UB, Mart\'{\i} i Franqu\`es 1, E08028 Barcelona, Spain
\and
Università degli Studi di Milano-Bicocca, piazza dell’Ateneo Nuovo 1, I-20126 Milano (MI), Italy
\and
Serra H\'unter Fellow
\and
School of Physics and Astronomy, University of Leicester, University Road, Leicester, LE1 7RH, UK
\and
Center for Interdisciplinary Exploration and Research in Astrophysics and Department of Physics and Astronomy, \\ Northwestern University, 2145 Sheridan Road, Evanston, IL 60208-3112, USA
\and
School of Physics and Astronomy, University of Birmingham, Birmingham, B15 2TT, UK
\and
Department of Astrophysics/IMAPP, Radboud University, PO Box 9010,
6500 GL, The Netherlands
\and
Department of Physics, University of Warwick,
Coventry, CV4 7AL, UK
}
\date{Received \dots; accepted \dots}

\abstract
{A total of four long duration Gamma-Ray Bursts (GRBs) have been confirmed at very high energy ($\ge$100\,GeV) with large significance hitherto, and any possible peculiarities of these bursts will become clearer as the number of detected events increases. Multi-wavelength follow-up campaigns are required to extract information on the physical conditions within the jets that lead to the very high energy counterpart, hence they are crucial to unveil the properties of this class of bursts.}
{GRB~201015A is a long-duration GRB detected using the MAGIC telescopes from $\sim$40\,s after the burst. If confirmed, this would be the fifth and least luminous GRB ever detected at this energies. The goal of this work is to constrain the global and microphysical parameters of its afterglow phase, and discuss the main properties of this burst in a broader context.}
{Since the radio band, together with frequent optical and X-rays observations, proved to be a fundamental tool to overcome the degeneracy in the afterglow modelling, we performed a radio follow-up of GRB~201015A over twelve different epochs, from 1.4 days (2020 October 17) to 117 days (2021 February 9) post-burst, with the Karl G. Jansky Very Large Array, e-MERLIN and the European VLBI Network. We include optical and X-rays observations, performed with the Multiple Mirror Telescope and the \emph{Chandra} X-ray Observatory respectively, together with publicly available data, in order to build the multi-wavelength light curves and to compare them with the standard fireball model.}
{We detected a point-like transient, consistent with the position of GRB~201015A until 23 and 47 days post-burst at 1.5 and 5\,GHz, respectively. No emission was detected in subsequent radio observations. The source was detected also in both optical (1.4 and 2.2 days post-burst) and X-ray (8.4 and 13.6 days post-burst) observations.}
{The multi-wavelength afterglow light curves can be explained with the standard model for a GRB seen on-axis, which expands and decelerates into a medium with a homogeneous density. A circumburst medium with a wind-like profile is disfavoured. Notwithstanding the high resolution provided by the VLBI, we could not pinpoint any expansion or centroid displacement of the outflow. If the GRB is seen at the viewing angle $\theta$ which maximises the apparent velocity $\beta_{app}$, i.e. $\theta \sim \beta_{app}^{-1}$, we estimate that the Lorentz factor for the possible proper motion is $\Gamma_{\alpha} \le$ 40 in right ascension and $\Gamma_{\delta} \le$ 61 in declination. On the other hand, if the GRB is seen on-axis, the size of the afterglow is $\le$5\,pc and $\le$16\,pc at 25 and 47 days. Finally, the early peak in the optical light curve suggests the presence of a reverse shock component before 0.01 days from the burst.}

\titlerunning{VLBI Observations of GRB 201015A}
\maketitle

\section{Introduction}
Long duration Gamma-Ray Bursts (GRBs) are extremely powerful flashes that generally last > 2s (\citealt{Mazets1981}, \citealt{Norris1984}, \citealt{Kouveliotou1993}) and whose prompt emission is detected mainly in the $\gamma$- and X-ray domains. They are thought to signpost the catastrophic death of a massive star (see e.g. \citealt{Galama1998}, \citealt{Piran2004}, \citealt{Kumar2015}), which has previously expelled its hydrogen envelope into the surrounding medium \citep{Woosley2006}, and the subsequent formation of a spinning, stellar mass black hole (BH; \citealt{Woosley1993}, \citealt{Paczynski1998}) or neutron star (NS; \citealt{Usov1992}). This newborn central engine may power and launch relativistic jets of ejected matter within which internal shocks \citep{Rees1994} or magnetic reconnections \citep{Drenkhahn2002} can convert a fraction of the bulk kinetic energy into the observed short-lived $\gamma$-ray radiation \citep{Meszaros2002}.
These expanding jets interact with the circumburst medium (\citealt{Rees1992}, \citealt{Meszaros1993}), triggering both a forward and a reverse shock (hereafter RS; \citealt{Meszaros2002}). The electrons at the shock fronts are accelerated to relativistic energies, producing a long-lived afterglow through synchrotron emission, which can be observed from high energies (GeV) through X-rays, optical and near-infrared down to the radio bands (see, e.g., \citealt{Gehrels2009}, \citealt{Kouveliotou2012}).

The radio light curve is fundamental for the afterglow modelling: together with frequent optical and X-ray observations, it helps us to better constrain the multi-dimensional parameter space and to distinguish between different scenarios, providing relevant information to understand the progenitor's nature and the GRB origin. Nevertheless, the detection rate of GRBs observed in the radio band is only $\sim$30\%, and an even smaller number of events has multi-epoch observations \citep{Chandra2012}. In events where radio emission has been detected, it can be observed for months or even years after the burst \citep{Piran2004}. It provides a powerful tool to better constrain not just the internal jet physics but also the geometry and physical evolution of the jet. Evidence of scintillation has helped study the expansion velocity of the outflow \citep{Frail1997}, achromatic light curve behaviour can inform on jet opening angle (``jet breaks") and constrain the transition from the relativistic to the non-relativistic expansion \citep{Frail2004} which can be used to infer the total kinetic energy by performing radio calorimetry (see, e.g., \citealt{Berger2004, Frail2005}). For the nearest events the high angular resolution provided by Very Long Baseline Interferometry (VLBI) has been proved to be complementary to studying the afterglow; it is a unique tool to measure the expansion \citep{Taylor2004} and the centroid displacement \citep{Mooley2018} of the outflow, to constrain its size \citep{Ghirlanda2019} and to distinguish the proper compact afterglow emission from contaminating components within the host galaxy. 

At the opposite end of the spectrum, only four GRBs have a bona fide detection in the Very High Energy (VHE; $\ge$100 GeV) range at either early epochs, i.e. GRB~190114C (300 GeV--1 TeV; \citealt{Magic2019}) and GRB~201216C (\citealt{Blanch2020b}), or at later times deep in the afterglow phase, namely GRB~180720B (100--400 GeV; \citealt{Abdalla2019}) and GRB~190829A (180 GeV--3.3 TeV; \citealt{HESS2021}). Studying this emission component allows to constrain the physical properties of the emitting region and/or of the shocked accelerated particles, and the most natural interpretation for this VHE emission is the Inverse Compton scattering of synchrotron photons, i.e. the Synchrotron Self-Compton (SSC). Based on the very few events detected so far, it seems that the VHE emission characterises both very energetic, such as GRB~180720B and GRB~190114C, and low energy events, such as 190829A, but any possible peculiarities of VHE detected bursts will become clearer as the sample of events increases. However, multi-wavelength follow up of these events has proved a fundamental tool to test the afterglow emission model: for GRB~190829A, e.g., the VHE emission detected by the H.E.S.S. telescopes has been firstly interpreted as synchrotron emission \citep{HESS2021}, while multi-wavelength follow up studies agree for an SSC emission origin \citep{Salafia2021, Zhang2021, Fraija2021}.

GRB~201015A was discovered on 2020 October 15 at 22:50:13 UT as a multi-peaked 10-s-duration GRB by the Neil Gehrels {\em Swift} Burst Alert Telescope (hereafter {\em Swift}/BAT) \citep{D'Elia2020a}. Subsequent observations reported the presence of an associated transient in the optical (\citealt{Lipunov2020a}; \citealt{Lipunov2020b}; \citealt{Malesani2020}; \citealt{Ackley2020}; \citealt{Hu2020}; \citealt{deUgartePostigo2020}; \citealt{Zhu2020a}; \citealt{Belkin2020a}; \citealt{Jelinek2020}; \citealt{Belkin2020b}; \citealt{Grossan2020}; \citealt{Rastinejad2020}; \citealt{Zhu2020b}; \citealt{Kumar2020a}; \citealt{Moskvitin2020}; \citealt{Pozanenko2020}), X-rays (\citealt{Kennea2020}; \citealt{Fletcher2020}; \citealt{Gompertz2020}; \citealt{D'Elia2020b}), UV \citep{Marshall2020} and radio \citep{Fong2020} bands. Remarkably, GRB~201015A was observed by the Major Atmospheric Gamma-ray Imaging Cherenkov (MAGIC) telescopes about 40\,s after the {\em Swift} trigger and a hint of a VHE counterpart with a significance $\ge$3.5$\sigma$ was reported from preliminary analyses \citep{Blanch2020a, Suda2021}. With the {\em Fermi} Gamma-ray Burst Monitor (GBM) spectrum, \cite{Minaev2020} suggested that this burst is consistent with the $E_\text{peak}-E_\text{iso}$ Amati relation \citep{Amati2002} for long-duration GRBs, with an isotropic equivalent energy of $E_\text{iso} \simeq (1.1 \pm 0.2)\times10^{50}$\,erg. If confirmed, this would be the fifth and least luminous GRB ever detected in this band.

Optical spectroscopy in the 3700--7800~\si{\angstrom} range revealed a redshift for the source of $\sim0.426$ \citep{deUgartePostigo2020, Izzo2020}; hitherto all the GRBs that have been detected at VHE have relatively low redshifts, i.e. 0.654, 0.425, 0.0785, and 1.1 for GRB~180720B, GRB~190114C, GRB~190829A and GRB~201216C respectively \citep{Vreeswijk2018, Selsing2019, Valeev2019, Vielfaure2020}, and their isotropic equivalent energies span three orders of magnitude \citep{Rhodes2020a}.

In this paper we present a multi-wavelength follow-up campaign of GRB~201015A performed with the Karl G. Jansky Very Large Array (VLA), the enhanced Multi Element Remotely Linked Interferometer Network (e-MERLIN), the European VLBI Network (EVN), the Multiple Mirror Telescope (MMT) and the \emph{Chandra} X-ray Observatory (\emph{Chandra}). The observations are presented in Section \ref{sec:obs}, while the results are shown in Section \ref{sec:results}. We exploit the standard model for GRB afterglows in Section \ref{sec:modeling} to explain the multi-wavelength observations and we compare our results for GRB~201015A with previous GRBs in Section \ref{sec:discussion}. We conclude with a brief summary in Section \ref{sec:conclusions}. Throughout the paper we assume a standard $\Lambda$-CDM cosmology with $H_{0} = 69.32$\,km\,Mpc$^{-1}$\,s$^{-1}$, $\Omega_{\rm m}=0.286$ and $\Omega_{\Lambda}=0.714$ \citep{Hinshaw2013}. With this cosmology, 1$^{\prime\prime}$ corresponds to roughly 5.6\,kpc at $z = 0.426$.

\section{Observations}
\label{sec:obs}
\subsection{VLA Observations at 6 GHz}
Observations with the VLA were performed 1.41 days post-burst (PI: Fong; project code: 19B-217) at a central frequency of 5.7\,GHz with a bandwidth of 1.6\,GHz (C-band). The target and the phase calibrator J2355+4950 were observed in 8 minutes cycles, with 7 minutes on the former and 1 minute on the latter, respectively. The distance between the target and the phase calibrator is about 4.5$^\circ$. Finally, 3C147 was used as bandpass and flux calibrator. Data were calibrated using the \textsc{casa} pipeline and they were subsequently imaged with the \texttt{tclean} task in \textsc{casa} (Version 5.1.1., \citealt{McMullin2007}).

\subsection{e-MERLIN Observations at 1.5 GHz}
We started observing at 1.5\,GHz with e-MERLIN 20 days post burst (2020 November 4; PI: Rhodes, project code: DD10003) with two further observations 23 (2020 November 7) and 101 (2021 January 24) days post-burst. The observations were made at a central frequency of 1.51\,GHz with a bandwidth of 512\,MHz (L-band). For each epoch, the target and phase calibrator, J2353$+$5518, were observed in 10 minute cycles, with 7 minutes on the former and 3 on the latter. The distance between the phase calibrator and the target is about 3$^\circ$. Each observation ended with scans of the flux (J1331+3030) and bandpass calibrators (1407+2827). The data were reduced using the custom e-MERLIN pipeline\footnote{\url{https://github.com/e-merlin/eMERLIN_CASA_pipeline}}. The calibrated measurement sets were imaged in \textsc{casa} (Version 4.7). 

\subsection{e-MERLIN Observations at 5 GHz}
Observations at 5\,GHz with e-MERLIN were performed 21 (2020 November 5), 24 (November 8), 60 (December 14), 85 (2021 January 8) and 100 (January 23) days post-burst (PI: Giroletti; project code: DD10004). All epochs but December 14 were centred between 4.50--5.01\,GHz
(C-band) with a bandwidth of 512\,MHz divided in four spectral windows of 128 MHz each. For the 14$^{\mathrm{th}}$ of December the frequency range was within 6.55--7.06\,GHz (C-band). Data were first pre-processed with the \textsc{casa} e-MERLIN pipeline using J1407+2827 as bandpass calibrator and J1331+3030 as flux calibrator. Two phase calibrators were used: J2353+5518, a fainter one on a rapid cycle, and J2322+5057, a brighter one used less frequently (once per hour), to correct for both short and long term atmospheric effects. All epochs were observed in 8 minutes cycles, with 6 minutes on the target and 2 minutes on J2353+5518.

On November 5 an electronic problem occurred and the Defford antenna missed the bandpass and flux calibrators; consequently, the pipeline automatically flagged out this antenna, with a considerable data loss. To recover it we performed a further calibration of this epoch: we built a model for J0319+4130 using the pipeline results first, and we subsequently calibrated the data manually using the J0319+4130 model as bandpass and flux calibrator, improving the final image output. After the calibration, we cleaned the dirty image with the \texttt{tclean} task in \textsc{casa} (Version 5.1.1.). 

On November 8, the Knockin antenna lost one polarisation channel, and an improved image was achieved using only J2322+5057 for the phase calibration, which is about 3.3$^\circ$ far from the target source.

\subsection{EVN Observations at 5 GHz}
EVN observations at 5\,GHz were performed 25 (2020 November 9), 47 (December 1), and 117 (2021  February 9) days post-burst (PI: Marcote; project code: RM016). The first epoch (2020 November 9) was conducted at a maximum bitrate of 4\,Gbps per station, dividing the full band upon correlation in 16 spectral windows of 32\,MHz and 64 frequency channels each, covering the frequency range of 4.57--5.11\,GHz (C-band). The other two following epochs were conducted at a lower rate of 2\,Gbps, resulting in eight spectral windows of 32\,MHz and 64 frequency channels each, covering the frequency range of 4.77--5.05\,GHz.
All observations were correlated in real time (e-EVN operational mode) at JIVE (The Netherlands) using the SFXC software correlator \citep{Keimpema2015}.

The following sources were used as fringe finders and/or bandpass calibrators among the different epochs: BL~LAC, J0854+2006, 3C~84, J0555+3948, and J0102+5824. The same phase calibrator as in the e-MERLIN observations was used: J2353+5518, in a phase-referencing cycle of 4.5 minutes on the target source and 1.5 minutes on the phase calibrator. The source J2347+5142 was observed as check source to account for possible phase-referencing losses.

The EVN data were reduced using {\tt AIPS}\footnote{The Astronomical Image Processing System ({\tt AIPS}) is a software package produced and maintained by the National Radio Astronomy Observatory (NRAO).} \citep{Greisen2003} and {\tt Difmap} \citep{Shepherd1994} following standard procedures. {\em A-priori} amplitude calibration was performed using the known gain curves and system temperature measurements recorded individually on each station during the observation.
We manually flagged data affected by radio frequency interference (RFI) and then we fringe-fitted and bandpass-calibrated the data using the fringe finders and the phase calibrator. We imaged and self-calibrated the phase calibrator in {\tt Difmap} to improve the final calibration of the data. We used the same model of the phase calibrator, obtained from the 2020 December 1 epoch, to improve the calibration of all epochs. We note that we chose this epoch because it produced the most reliable image of J2353+5518 in terms of amplitude scales at all baseline lengths (including the short spacing given by the e-MERLIN stations). No apparent changes in the calibrator were observed among these three observations. The obtained solutions were then transferred to the target scans, which were subsequently imaged for each epoch. The check source J2347+5142 was also imaged and self-calibrated, confirming that no significant ($\lesssim 10$--$20\%$) losses were present in the obtained amplitudes due to the phase-referencing technique. We stress that the Shanghai 65\,m Radio Telescope (Tianma) and the Nanshan 25\,m Radio Telescope (Urumqi) only participated in the first observation, and since they provided the longest baselines the resolution for the other two epochs decreases significantly (see Table~\ref{tab:obs}).

\subsection{Optical Observations and Public Data}

At 1.4, 2.2 and 4.3 days post-burst, we observed the position of the afterglow in the $i$- and $z$-bands with the Binospec instrument mounted on the 6.5m MMT (PI: Fong; project code: 2020c-UAO-G204-20B). We reduced our images using a custom Python pipeline\footnote{\url{https://github.com/CIERA-Transients/Imaging_pipelines/}} and registered the images to the USNO-B1 catalogue \citep{USNO-B1} using standard IRAF tasks \citep{Tody93}. In the first two epochs, we clearly detected an uncatalogued source in both bands that did not appear in our deep image at 4.3 days post-burst. To remove any contamination from the nearby galaxy, we performed image subtractions between the first two epochs and the final epoch using \texttt{HOTPANTS} \citep{Becker15}. We then calibrated the images to the PanSTARRS Data Release 2 catalogue \citep{Chambers+16} and performed aperture photometry on the image subtractions with the IRAF/\texttt{phot} task.

We gathered additional optical information from the public GCN Circulars Archive, and the detected emission was de-absorbed with the \texttt{dust\_extinction} Python package\footnote{\url{https://dust-extinction.readthedocs.io/en/stable/} }, using a Galactic extinction $A_{v}$ = 0.93 \citep{Schlafly2011}.

\subsection{X-ray Observations and Public data}
We obtained the {\em Swift} X-ray Telescope (XRT) unabsorbed flux light curve integrated in the 0.3--10\,keV energy range from the \textsc{Swift Burst Analyzer}\footnote{\url{https://www.swift.ac.uk/burst_analyser/01000452/}} provided by the UK {\em Swift} Science Data Centre at the University of Leicester (UKSSDC, \citealt{Evans2007,Evans2009}).

Moreover, we obtained two epochs of \emph{Chandra} observations with the Advanced CCD Imaging Spectrometer (ACIS) in very faint mode (PI: Gompertz; project code: 22400511). Exposures were centred around $8.4$ and $13.6$ days after trigger, with exposure times of 30\,ks and 45\,ks respectively. The data were analysed using {\sc CIAO v4.14} and {\sc XSPEC v12.11.1}, following the \emph{Chandra} X-ray Observatory science threads\footnote{\url{https://cxc.harvard.edu/ciao/}}.

\section{Results}
\label{sec:results}
\subsection{Radio}
A point-like source was clearly visible with the VLA 1.4 days post-burst with a peak brightness of 132$\pm$8\,$\mathrm{\upmu Jy\ beam^{-1}}$, where the uncertainty includes the r.m.s noise and a 5\% calibration error added in quadrature. The r.m.s noise uncertainty is 5\,$\mathrm{\upmu Jy\ beam^{-1}}$, and therefore the detection has a significance of 26$\sigma$ confidence. The source was found at a position (J2000) $\alpha = 23^{\rm h}37^{\rm m}16.403^{\rm s}$, $\delta = 53^\circ24^\prime56.39^{\prime\prime}$, with an uncertainty of $0.14^{\prime\prime}$ ($1/10$ of the beam size, \citealt{Taylor1999}). Thanks to the wide bandwidth and high signal to noise ratio, we could split the data in four spectral windows in order to estimate the spectral index $\beta$, where the flux density is $F \propto \nu^{\beta}$. We found $\beta \simeq 2.5$. To further improve this estimate, we produced a spectral map with the \texttt{tclean} task in \textsc{casa} by setting nterms=2 and deconvolver=`mtmfs'. We found $\beta = 2.3\pm0.1$ at the peak of the target emission. We attribute the emission to the afterglow of GRB~201015A. Finally, we divided the 1 hour long observation in two intervals of equal duration and determined the peak brightness in each one, which turned out to be 126$\pm$9\,$\mathrm{\upmu Jy\ beam^{-1}}$ and 144$\pm$10\,$\mathrm{\upmu Jy\ beam^{-1}}$ respectively (see Figure \ref{fig:MWL_LC_ISM}, blue stars).

The resulting images from the first and second e-MERLIN epoch at 1.5\,GHz showed a point source with a peak brightness of 213$\pm$40\,$\mathrm{\upmu Jy\ beam^{-1}}$ and 261$\pm$48\,$\mathrm{\upmu Jy\ beam^{-1}}$, where the quoted uncertainty includes the r.m.s noise and a 10\% calibration error added in quadrature, at the position (J2000) $\alpha =23^{\rm h}37^{\rm m}16.423^{\rm s}$, $\delta = +53^{\circ}24^{\prime}56.43^{\prime\prime}$. The r.m.s. noise uncertainties are 34\,$\mathrm{\upmu Jy\ beam^{-1}}$ and 40\,$\mathrm{\upmu Jy\ beam^{-1}}$, hence the detections have a significance of 6.2 and 6.5$\sigma$ confidence, respectively. The uncertainty on the position, which was computed as the ratio between the beam size and the signal to noise ratio \citep{Taylor1999}, is 0.03$^{\prime\prime}$. Unfortunately, the observation at 101 days was heavily affected by RFI and as a result we obtained a 5$\sigma$ upper limit of 285\,$\mathrm{\upmu Jy\ beam^{-1}}$. 
Data are shown in Figure~\ref{fig:MWL_LC_ISM} as gold squares.

\begin{figure}
\centering
\includegraphics[width=0.9\columnwidth]{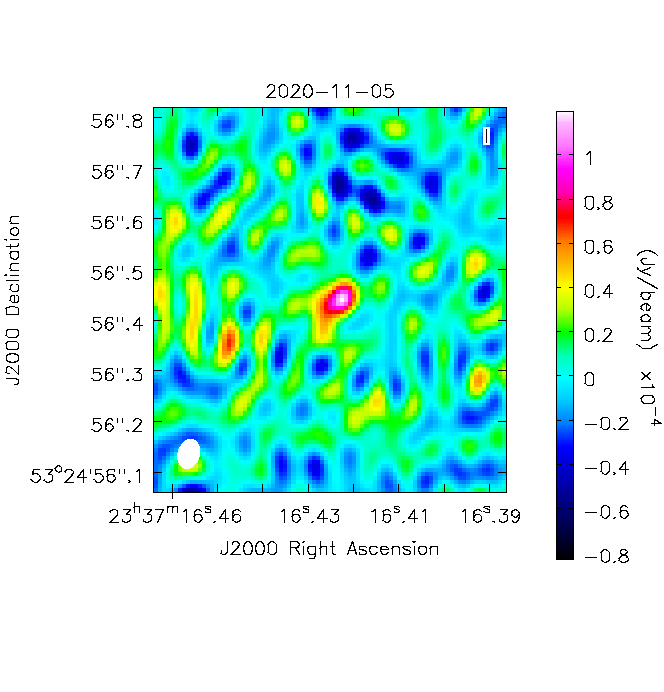}
\caption{e-MERLIN detection on 2020 November 5. The synthesised beam is shown on the lower left.}
\label{fig:image}
\end{figure}

At 5\,GHz a point-like transient was clearly detected with e-MERLIN on November 5 (Figure \ref{fig:image}) at the position (J2000) of $\alpha = 23^{\rm h}37^{\rm m}16.422^{\rm s}$, $\delta = 53^{\circ}24^{\prime}56.44^{\prime\prime}$. The uncertainty on the position is 0.01$^{\prime\prime}$. The point-like source was also detected on November 8 at the position (J2000) $\alpha = 23^{\rm h}37^{\rm m}16.419^{\rm s}$, $\delta = 53^\circ24^\prime56.33^{\prime\prime}$. The uncertainty on the position is 0.02$^{\prime\prime}$. 
Albeit both positions are in agreement with the coordinates provided by the VLA, we note that they are not consistent with each other at 3$\sigma$ confidence level. We ascribe the offset in the position to the phase calibration of the second epoch: if the phase calibrator is observed less frequently (i.e. once per hour), it may not be able to trace perfectly, and therefore correct, the short-term atmospheric effects. Nevertheless, we were not able to improve the phase calibration further. The measured peak brightness is 107$\pm$20\,$\mathrm{\upmu Jy\ beam^{-1}}$ and 116$\pm$28\,$\mathrm{\upmu Jy\ beam^{-1}}$ for November 5 and 8 respectively, where the quoted uncertainty includes the r.m.s noise uncertainty and a 10\% calibration error added in quadrature. The r.m.s. noise uncertainties are 17\,$\mathrm{\upmu Jy\ beam^{-1}}$ and 26\,$\mathrm{\upmu Jy\ beam^{-1}}$, hence the detections have a significance of 6.3 and 4.5$\sigma$ confidence, respectively. On December 14, January 8 and 23 no source was detected. The r.m.s. noise is 43, 19 and 16 $\mathrm{\upmu Jy\ beam^{-1}}$, respectively. Data are shown in Figure \ref{fig:MWL_LC_ISM} as blue dots.

GRB~201015A was detected as a point-like source also in the first two epochs with EVN at 5\,GHz (25 and 47 days after the burst) at a consistent (J2000) position of $\alpha = 23^{\rm h}37^{\rm m}16.42232^{\rm s} \pm 0.2\,\mathrm{mas},\ \delta = 53^\circ24^\prime56.439{\bf 2}^{\prime\prime} \pm 0.3\,\mathrm{mas}$. The quoted uncertainties include the statistical uncertainties (0.05 and 0.12 mas for $\alpha$ and $\delta$, respectively), the uncertainties in the absolute International Celestial Reference Frame position of the phase calibrator (0.11~mas), and check source (0.15~mas; \citealt{beasley2002,gordon2016}), and the estimated uncertainties from the phase-referencing technique (0.13 and 0.2~mas; \citealt{pradel2006}) added in quadrature.

The derived peak brightness measurements are $85\pm13$\,$\mathrm{\upmu Jy\ beam^{-1}}$ and $73\pm12$\,$\mathrm{\upmu Jy\ beam^{-1}}$ respectively, where the errors comprise both the r.m.s noise uncertainty and a 10\% calibration error, added in quadrature. The r.m.s. noise uncertainties are 9\,$\mathrm{\upmu Jy\ beam^{-1}}$ and 10\,$\mathrm{\upmu Jy\ beam^{-1}}$, hence the detections have a significance of 9.4 and 7.3$\sigma$ confidence, respectively. No significant emission above the 3$\sigma$ r.m.s. level ($\sigma =$13\,$\mathrm{\upmu Jy\ beam^{-1}}$) was reported in the third epoch. Data are shown in Figure \ref{fig:MWL_LC_ISM} as blue squares.

The upper limits for the flux densities in the radio band were taken with 3$\sigma$ confidence level. The full list of radio observations is given in Table~\ref{tab:obs}.

\begin{table*}[t]
\resizebox{18cm}{!}{
\centering
\begin{tabular}{ccccccccc}
\toprule
Date    &UTC    &T-T$_0$    &T$_s$  &$\nu$ &Peak Brightness   &r.m.s.  &Array  &Beam size\\
        &[hh:mm]    &[days]   &[hour]      &[GHz]   &[$\upmu$Jy/beam] &[$\upmu$Jy/beam]  &    &\\
\midrule
2020/10/17 &8:58 -- 9:38  &1.4 &0.7 &4.23 -- 7.10   &132   &5    &VLA    &1.70$^{\prime\prime} \times$ 1.14$^{\prime\prime}$\\
2020/11/04 &21:25 -- 06:30  &20 &9.2 &1.25 -- 1.76   &213   &34    &e-MERLIN    &0.18$^{\prime\prime} \times$ 0.12$^{\prime\prime}$\\
2020/11/05 &20:35 -- 14:00  &21 &6.1  &4.50 -- 5.01 &107 &17    &e-MERLIN    &0.06$^{\prime\prime} \times$ 0.04$^{\prime\prime}$\\
2020/11/07 &22:00 -- 11:40  &23 &14  &1.25 -- 1.76   &261   &40    &e-MERLIN    &0.19$^{\prime\prime} \times$ 0.12$^{\prime\prime}$\\
2020/11/08 &23:30 -- 08:30  &24 &3.9  &4.50 -- 5.01 &116 &26    &e-MERLIN    &0.06$^{\prime\prime} \times$ 0.04$^{\prime\prime}$\\
2020/11/09 &13:00 -- 23:00  &25 &4.2 &4.57 -- 5.11  &85 &9    &EVN    &1.8\,mas $\times$ 0.9\,mas\\
2020/12/01 &13:00 -- 23:00  &47 &4.4 &4.77 -- 5.05 &73 &10 &EVN    &3.4\,mas $\times$ 2.8\,mas\\
2020/12/14 &09:18 -- 12:43  &60 &1.4  &6.55 -- 7.06    &- &43    &e-MERLIN    &0.12$^{\prime\prime} \times$ 0.07$^{\prime\prime}$\\
2021/01/08 &12:34 -- 03:10  &85 &6.9 &4.50 -- 5.01  &- &19    &e-MERLIN    &0.04$^{\prime\prime} \times$ 0.04$^{\prime\prime}$\\
2021/01/23 &17:35 -- 08:55  &100 &8.9 &4.50 -- 5.01  &- &16    &e-MERLIN    &0.07$^{\prime\prime} \times$ 0.03$^{\prime\prime}$\\
2021/01/24 &11:00 -- 01:20  &101 &14   &1.25 -- 1.76   &-   &57    &e-MERLIN    &0.17$^{\prime\prime} \times$ 0.14$^{\prime\prime}$\\
2021/02/09 &13:00 -- 18:00 \& 06:00 -- 11:00  &117 &5.0 &4.77 -- 5.05 &- &13   &EVN    &3.1\,mas $\times$ 3.6\,mas\\
\bottomrule
\end{tabular}}
\caption[]{Radio observations performed with the VLA, e-MERLIN and EVN in the L- and C-bands. T-T$_0$ is the total time from the GRB trigger to half of the observation, while T$_s$ is the total time on source. The 1$\sigma$ r.m.s. noise shown does not include the systematic flux density uncertainty (which we have considered as 5\% for the VLA and 10\% for e-MERLIN and EVN throughout the work).}
\label{tab:obs}
\end{table*}

\subsection{Optical}
At 1.4 and 2.2 days post-burst, we clearly detected the optical afterglow in both $i$- and $z$-band at $\alpha =23^{\rm h}37^{\rm m}16.43^{\rm s}$, $\delta = +53^{\circ}24^{\prime}56.6^{\prime\prime}$ (J2000; uncertainty = 0.2$^{\prime\prime}$). In addition, we detected the host galaxy at $\alpha =23^{\rm h}37^{\rm m}16.48^{\rm s}$, $\delta = +53^{\circ}24^{\prime}54.6^{\prime\prime}$ (J2000; uncertainty = 0.2$^{\prime\prime}$).

The optical light curve is shown in Figure \ref{fig:MWL_LC_ISM}: $g$-band data from \citealt{Belkin2020b} (green hexagons in Figure \ref{fig:MWL_LC_ISM}), \citealt{Grossan2020} (green dots) and \citealt{Ackley2020} (green circles); $r$-band data from \citealt{Belkin2020a} (red pentagons), \citealt{Belkin2020b} (red stars), \citealt{Zhu2020a, Zhu2020b} (red hexagons), \citealt{Moskvitin2020} (red diamonds), \citealt{Grossan2020} (thin red diamonds), \citealt{Kumar2020b} (red plus), \citealt{Pozanenko2020} (red circles); $i$-band data from \citealt{Grossan2020} (purple squares) and our MMT/Binospec observations (purple circles); our $z$-band MMT/Binospec observations (brown circles).

The emission peaked between 200--300~s after the GRB trigger, reaching a maximum of R$\sim$16.5 mag (\citealt{Jelinek2020}; \citealt{Zhu2020a}). Between 0.1 and 3 days our light curve follows a power law F$(t) \propto t^{-0.84\pm0.06}$, which is consistent with previous results in the GCNs \citep{Pozanenko2020}. Remarkably, a type Ic-BL supernova (SN) contribution can be seen between 3 and 20 days after the burst (\citealt{Pozanenko2020}; \citealt{Rossi2020}), which corroborates the long-duration nature of this burst. 

\subsection{X-rays}
The \emph{Swift}/XRT light curve was further analysed by splitting the last two observations in four time intervals. We retrieved the XRT spectral files from the online archive\footnote{https://www.swift.ac.uk/xrt\_spectra} and analyse them with the public software {\sc XSPEC v12.10.1f}, assuming a simple power-law model.
The \texttt{tbabs} model for the Galactic absorption and the \texttt{ztbabs} model for the host galaxy absorption, adopting the source redshift $z=0.426$, are used in the fitting procedure. The absorption parameters are fixed to the values reported by the {\em Swift} website for this burst, namely $N_{\rm H, gal}=3.6\times10^{21}$\,atoms\,cm$^{-2}$ \citep{Kalberla2005,Willingale2013} and $N_{\rm H, intr}=5\times10^{21}$\,atoms\,cm$^{-2}$.

Leaving the normalisation and the photon index of the power-law free to vary, we find integrated fluxes consistent with the ones reported by the {\em Swift} website.

From our two epochs of \emph{Chandra} observations we find $0.5$ -- $7$\,keV source count rates of $(4.07\pm0.38)\times10^{-3}$\,cts/s and $(3.11\pm0.29)\times10^{-3}$\,cts/s. In a combined spectral fit of both \emph{Chandra} epochs and the late ($> 10$\,days) XRT observations, the data are well modelled (cstat/dof $= 600/1808$) by an absorbed power law of the form {\sc powerlaw*tbabs*ztbabs} \citep{Wilms2000} with a photon index of $\Gamma = 2.10 \pm 0.13$. The intrinsic absorption column is fixed to $N_{\rm H, intr} = 5 \times 10^{21}$\,atoms\,cm$^{-2}$ at $z = 0.426$ over the Galactic value of $N_{\rm H, gal} = 3.6\times 10^{21}$\,atoms\,cm$^{-2}$ \citep{Kalberla2005,Willingale2013} to match those reported on the UKSSDC. From this, we derived unabsorbed $0.3$--$10$\,keV fluxes of $(1.26 \pm 0.05)\times 10^{-13}$\,erg\,cm$^{-2}$\,s$^{-1}$ at $8.4$ days and $(1.10 \pm 0.04) \times 10^{-13}$\,\,erg\,cm$^{-2}$\,s$^{-1}$ at $13.6$ days.

\begin{figure*}
\centering
\includegraphics[width=0.9\textwidth]{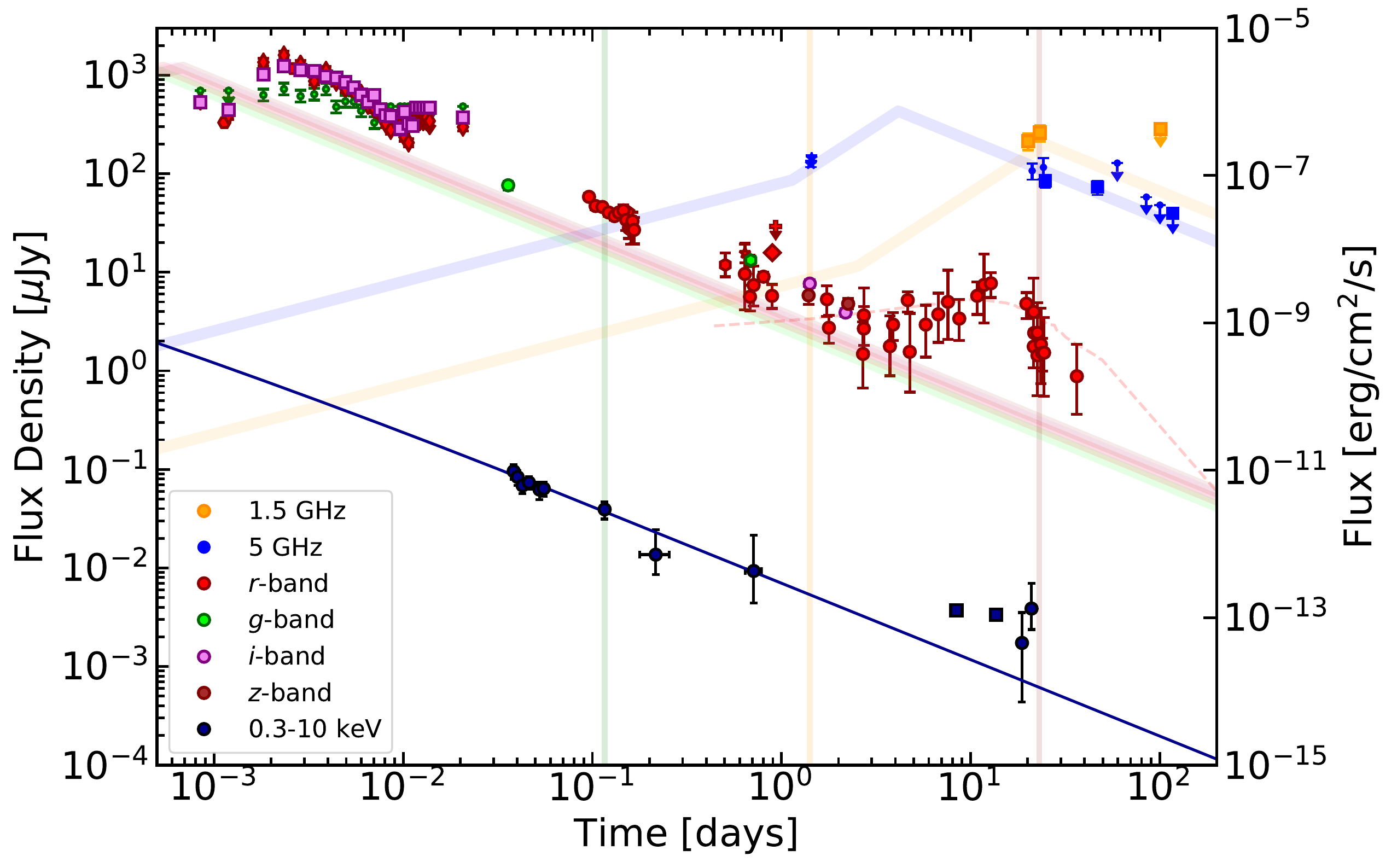}
\caption{Multi-wavelength afterglow light curves (see Section \ref{sec:results}). For each band, the light curves predicted by the standard model with $\nu_{\rm sa}$ = 13\,GHz, $\nu_{\rm m}$ = 6\,GHz, $\nu_{\rm c}$ = 2$\times10^{7}$\,GHz, $F_{\rm m}$ = 800\,$\mathrm{\upmu Jy}$ and $p$ = 2.05 at 1 day for a homogeneous surrounding medium are shown: 1.5 GHz (orange), 5 GHz (blue), $r$, $g$, $i$ and $z$- bands (red, lime, violet and brown respectively), integrated X-ray light curve (dark blue). The green, orange and brown vertical lines pinpoint the epochs of the spectra at 0.12, 1.41 and 23 days respectively (see also Figure \ref{fig:spectra}). The dashed line shows a simple model for the SN contribution in the $r$-band (see Section \ref{subsec:add_comp}).}
\label{fig:MWL_LC_ISM}
\end{figure*}

The X-ray light curve is shown in Figure \ref{fig:MWL_LC_ISM}: the \emph{Swift}/XRT public data (dark blue circles) and our \emph{Chandra} observations (dark blue squares). For the \emph{Swift}/XRT light curve we included the results from the \textsc{Swift Burst Analyzer} up to $\sim$0.12 days, and from that epoch on we used our re-analysis of the last two observations. Our XRT analysis suggests that the light curve can be fitted with a power law with index $F^{-1.1\pm0.3}$ between 0.04 and 0.71 days post-burst, which is shallower but still consistent with the previous analysis from \cite{DAi2020}. However, the subsequent detections at 8.4 and 13.6 days with {\em Chandra} show a flux $\sim$6 and 8 times higher than expected from extrapolating the earlier XRT light curve respectively, and the increased flux is further confirmed by the late time ($\sim$20 days after the burst) {\emph Swift}/XRT follow-up \citep{D'Elia2020b}.
\begin{figure*}
\centering
\includegraphics[width=0.7\textwidth]{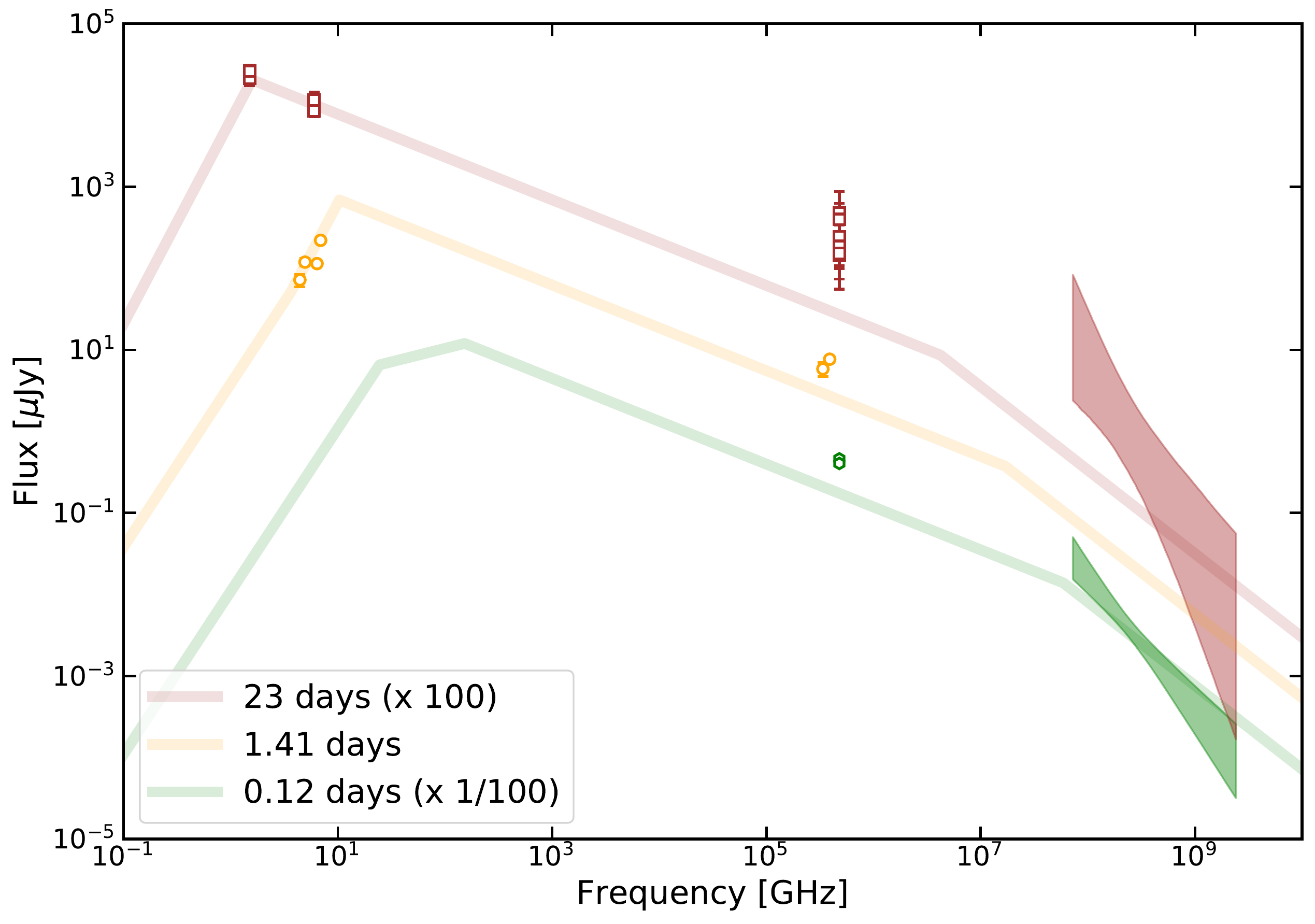}
\caption{Spectra at 0.12 (green), 1.41 (orange) and 23 (brown) days after the GRB onset for a homogeneous surrounding medium with $\nu_{\rm sa}$ = 13\,GHz, $\nu_{\rm m}$ = 6\,GHz, $\nu_{\rm c}$ = 2$\times10^{7}$\,GHz, $F_{\rm m}$ = 800\,$\mathrm{\upmu Jy}$ and $p$ = 2.05 at 1 day. \emph{Spectrum at 0.12 days}: optical observations from \citet{Pozanenko2020} and XRT butterfly plot. \emph{Spectrum at 1.41 days}: our VLA and MMT detections. \emph{Spectrum at 23 days}: our 1.5 and 5\,GHz observations, optical data from \citet{Pozanenko2020} and the XRT butterfly plot; the optical emission is dominated by the SN component.}
\label{fig:spectra}
\end{figure*}

\section{Broadband Modelling}
\label{sec:modeling}
The multi-wavelength afterglow synchrotron emission of a GRB seen on-axis can be studied through a standard model (see, e.g., \citealt{Granot2002}, \citealt{Zhang2004}). First, assuming that the flux density can be parameterised as $F\propto\nu^{\beta} t^{\,\alpha}$, the spectrum can be fitted with several power law segments, which join at specific break frequencies: (i) the self-absorption frequency $\nu_{\rm sa}$, (ii) the maximum frequency $\nu_{\rm m}$ and (iii) the cooling frequency $\nu_{\rm c}$. The other parameters needed to build the spectrum are (iv) the maximum flux density $F_{\rm m}$ and (v) the electron distribution index $p$. Once we have determined these quantities and their temporal evolution, the multi-wavelength light curves are constrained. For this work we use the relations provided by \citet{Granot2002}, and throughout the paper we consider two possible density profiles for the circumburst medium: a wind-like profile $\rho$ = $Ar^{-2}$, which is naturally expected if the progenitor is a massive star collapsing into a BH or a NS, and a homogeneous surrounding medium $\rho=\mathrm{const}$, which can be ascribed either to the canonical ISM or to a wind bubble shocked against the ISM \citep{Aksulu2021}. Hereafter we will use the term ISM for a homogeneous profile indistinctly.

We stress that in our modelling we do not include the description of the coasting phase, the contribution from the RS, nor the late time SN emission. As a matter of fact, a more sophisticated modelling that comprises the RS contribution would introduce more parameters; if frequent observations are available around the epoch at which the RS is supposed to prevail (at about 1 day in the radio band; see, e.g., \citealt{Rhodes2020a}) these parameters can be constrained. With only one detection in the C-band before 20 days post-burst, we could not constrain the parameters. Concerning the optical, the emission before 0.01 days shows a bump which could be due to a possible RS contribution, while after 3 days the SN emission becomes dominant (\citealt{Pozanenko2020, Rossi2020}), hence the prediction of the modelling should be considered only from about 0.01 to 3 days post-burst in this band.

To derive the modelling light curves, we perform a comparison of the simplified afterglow prescription with the available data, changing the aforementioned parameters to get as close as possible to the observed multi-wavelength light curves and to reproduce the afterglow spectrum at three sampling epochs, namely 0.12, 1.41 and 23 days after the GRB trigger (see Figure \ref{fig:spectra}).

\subsection{ISM profile}
For the ISM profile, we build the spectrum at at 0.12 days, with the optical $r$-band from \citet{Pozanenko2020} and the XRT detections, 1.41 days, with the VLA detection (see Section \ref{sec:results}) and our optical $i$ and $z$-band observations, and the spectrum at 23 days, with our radio detection at 1.5 and 5\,GHz, the optical $r$-band from \citet{Pozanenko2020} and the last XRT detection (Figure \ref{fig:spectra}). From the spectra and the multi-wavelength light curves we constrain the parameter space as follows.
First, from the spectral index $\beta=2.3\pm0.1$ derived with the VLA data we cannot discern whether the emission at 6\,GHz lies in the $\nu^{2}$ or $\nu^{5/2}$ portion of the spectrum at 1.41 days, and therefore we consider three different cases: at this epoch it could be that (i) 6\,GHz < $\nu_{\rm sa}$ < $\nu_{\rm m}$, (ii) $\nu_{\rm m}$ < 6\,GHz < $\nu_{\rm sa}$, (iii) 6\,GHz < $\nu_{\rm m} \leq \nu_{\rm sa}$. Moreover, at 23 days the spectral slope between 1.5\,GHz and 5\,GHz is reversed, i.e. the flux density is decreasing with the frequency, and hence we expect that $\nu_{\rm m}$ < $\nu_{\rm sa}$ < 1.5\,GHz. Finally, at 23 days the optical emission is dominated by the SN, hence we consider the optical detections as upper limits. To build the modelling light curves and spectra we derive the break frequencies, the $p$ value and the maximum flux density $F_{\rm m}$ at 1 day, in order to simplify the equations from \cite{Granot2002}.

\begin{itemize}
\item[(i)] If 6\,GHz < $\nu_{\rm sa}$ < $\nu_{\rm m}$, since $\nu_{\rm m}$ > 6\,GHz at 1.41 days, $\nu_{\rm m}\propto t^{-3/2}$ and $\nu_{\rm sa}$ is constant in time, to avoid $\nu_{\rm m}$ crossing $\nu_{\rm sa}$ before 1.41 days we shall impose $\nu_{\rm m}$ > 15\,GHz and $\nu_{\rm sa}$ > 9\,GHz at 1 day. However, once $\nu_{\rm m}$ crosses $\nu_{\rm sa}$, $\nu_{\rm sa} \propto t^{-(3p+2)/2(p+4)}$. Therefore at 1 day $\nu_{\rm sa}$ < 13\,GHz, otherwise at 23 days $\nu_{\rm sa}$ > 1.5\,GHz, and consequently $\nu_{\rm m}$ < 24\,GHz (otherwise it does not cross $\nu_{\rm sa}$ before 23 days). At 1 day the flux density at $\nu_{\rm m}$ is found to be 500\,$\mathrm{\upmu Jy}$ < $F_{\rm m}$ < 600\,$\mathrm{\upmu Jy}$: with a lower $F_{\rm m}$ we underestimate the emission at 5\,GHz observed with EVN, while with a higher flux we overestimate the e-MERLIN detections at the same frequency. With the slope of the optical light curve we can constrain the $p$ value: since the light curve shows a clear slope that can be described by a single power law between 0.01 and 3 days, $\nu_{\rm m}$ < optical < $\nu_{\rm c}$ and $F \propto t^{3(1-p)/4}$ in this regime. Finally, the X-ray integrated light curve allows us to further constrain $p$ and determine $\nu_{\rm c}$: for $\nu$ < $\nu_{\rm c}$ we have $F \propto t^{3(1-p)/4}$ while for $\nu$ > $\nu_{\rm c}$ we have $F \propto t^{(2-3p)/4}$; hence the sooner $\nu_{\rm c}$ crosses the X-ray band, the fainter the detected emission will be. To sum up, to reproduce both the spectra and the light curves we find that 9\,GHz < $\nu_{\rm sa}$ < 13\,GHz, 15\,GHz < $\nu_{\rm m}$ < 24\,GHz, 5$\times10^6$\,GHz < $\nu_{\rm c}$ < $10^8$\,GHz, 500\,$\mathrm{\upmu Jy}$ < $F_{\rm m}$ < 600\,$\mathrm{\upmu Jy}$ and 2.01 < $p$ < 2.10 at 1 day.

\item[(ii)] If $\nu_{\rm m}$ < 6\,GHz < $\nu_{\rm sa}$ at 1.41 days, since $\nu_{\rm sa}\propto t^{-(3p+2)/2(p+4)}$ we shall impose that $\nu_{\rm sa}$ > 10\,GHz at 1 day; moreover, $\nu_{\rm sa}$ < 18\,GHz at 1 day, otherwise at 23 days $\nu_{\rm sa}$ > 2\,GHz and our detections at 1.5\,GHz would lie in the $\nu^{5/2}$ portion of the spectrum and the emission at 5\,GHz would be overestimated. To reproduce the spectra and the light curves we find that the range for $\nu_{\rm sa}$ is further constrained to 13\,GHz < $\nu_{\rm sa}$ < 16\,GHz. Since at 1.41 days $\nu_{\rm m} \le$ 4\,GHz (otherwise the lowest end of the bandwidth of the VLA detection would be underestimated), at 1 day $\nu_{\rm m} \le$ 7\,GHz. Finally, with the same argument presented in case (i), we find that at 1 day 6$\times10^6$\,GHz < $\nu_{\rm c}$ < $10^8$\,GHz, 800\,$\mathrm{\upmu Jy}$ < $F_{\rm m}$ < 1\,mJy and 2.01 < $p$ < 2.20. We stress that in this case $F_{\rm m}$ refers to the flux density at $\nu_{\rm sa}$.

\item[(iii)] If 6\,GHz < $\nu_{\rm m} \leq \nu_{\rm sa}$ at 1.41 days we can have both 6\,GHz < $\nu_{\rm m}$ < $\nu_{\rm sa}$ and 6\,GHz < $\nu_{\rm sa}$ < $\nu_{\rm m}$ at 1 day. Considering both these sub-cases, since $\nu_{\rm m}\propto t^{-3/2}$, at 1 day $\nu_{\rm m}$ > 13\,GHz, otherwise at 1.41 days $\nu_{\rm m}$ < 8\,GHz and it would lie too close to the highest end of the bandwidth of the VLA detection to reproduce the spectrum; conversely, if at 1 day $\nu_{\rm m}$ > 18\,GHz we cannot reproduce the light curve in the C-band, i.e. the detections at 6\,GHz with the VLA are underestimated, while e-MERLIN and EVN observations are overestimated. Since at 1.41 days $\nu_{\rm sa} \geq \nu_{\rm m}$, we find that 13\,GHz < $\nu_{\rm sa}$ < 18\,GHz (for larger values we cannot reproduce the C-band light curve). Once again, with the same argument presented in case (i), we derived $5\times10^6$\,GHz < $\nu_{\rm c}$ < $2\times10^8$\,GHz, 630\,$\mathrm{\upmu Jy}$ < $F$ < 1\,mJy and 2.01 < $p$ < 2.20 at 1 day. In this case $F_{\rm m}$ refers to the flux density of $\nu_{\rm sa}$ or $\nu_{\rm m}$ for the two sub-cases respectively. We stress that the aforementioned ranges for the parameters are the superposition of the ranges derived for both the sub-cases.
\end{itemize}

In Table \ref{tab:params} we report our results for the parameter space at 1 day. The model light curves for the ISM profile are shown in Figure \ref{fig:MWL_LC_ISM} for $\nu_{\rm sa}$ = 13\,GHz, $\nu_{\rm m}$ = 6\,GHz, $\nu_{\rm c}$ = 2$\times$10$^{7}$\,GHz, $F_{\rm m}$ = 800\,$\mathrm{\upmu Jy}$ at 1 day, and an electron distribution index $p$ = 2.05. The 1.5\,GHz and the 5\,GHz light curve are displayed in orange and blue, respectively; the $r$, $g$, $i$ and $z$ bands are in red, lime, violet and brown respectively; the X-ray light curve is displayed in dark blue. Albeit this modelling provides a satisfactory description of the multi-wavelength light curves, the optical light curve contains the already discussed features in addition to the forward shock emission: before 0.01 days there is a bump which could be due to a possible RS contribution, while after three days the SN emission becomes dominant (\citealt{Pozanenko2020, Rossi2020}). 
\begin{table}[t]
\centering
\begin{tabular}{cc}
\toprule
Parameter    &Range\\
\midrule
$\nu_{\rm sa}$   &9 -- 18\,GHz\\
$\nu_{\rm m}$   &$\le$7\,GHz $\bigcup$ 13 -- 24\,GHz\\
$\nu_{\rm c}$   &5$\times10^{6}$ -- 2$\times10^{8}$\,GHz\\
$F_{\rm m}$   &0.5--1\,mJy\\
$p$   &2.01 -- 2.20\\
\bottomrule
\end{tabular}
\caption[]{Constraints on the model parameters at 1 day for a homogeneous circum-burst medium.}
\label{tab:params}
\end{table}

\subsection{Wind-like profile}
For the wind-like profile we first try to reproduce the optical and X-ray data, finding that $\nu_{\rm sa}$ = 1\,GHz, $\nu_{\rm m}$ = 30\,GHz, $\nu_{\rm c}$ = 2 $\times$10$^{7}$\,GHz, $F_{\rm m}$ = 200\,$\mathrm{\upmu Jy}$ at 1 day, and the electron distribution index $p$ = 2.01. Since this model conspicuously fails at reproducing the radio detections and the optical slope, we try to reproduce the radio light curve at 5\,GHz first, and we find that $\nu_{\rm sa}$ = 4\,GHz, $\nu_{\rm m} = 10^{3}$\,GHz, $\nu_{\rm c}$ = 2 $\times$10$^{7}$\,GHz, $F_{\rm m}$ = 600 $\mathrm{\upmu Jy}$ at 1 day, and the electron distribution index $p$ = 2.01. Neither of these models can reproduce the optical slope, and the latter fails at reproducing the X-ray emission. Different choices of the parameters in the wind-like scenario provide even poorer fits. We can therefore conclude that the modelling provided by the ISM provides the best agreement with the data, and we consider it hereafter. We shall note that this further corroborates the need of X-ray, optical and radio observations in order to break the degeneracy in the afterglow modelling, as with only two of them data can be misinterpreted.

\subsection{Intrinsic Host Galaxy Extinction}
\label{subsec:intrinsic_ext}
As pointed out by \citet{Kann2006}, the intrinsic, host galaxy extinction can be relevant in the optical/NIR. By changing the model parameters, we tried overestimating the optical emission and, from the discrepancy between the observed and the modelled optical flux densities, the contribution due to the intrinsic host galaxy absorption can be estimated. However, our modelling light curves and spectra cannot predict values for the flux density that are larger than those observed in the optical data. Moreover, by changing the maximum flux density and the p-value, we cannot reproduce anymore the observed light curves in the radio band. 
As our modelling light curve already underestimates the afterglow optical emission (see, e.g., Figure \ref{fig:spectra}), by adding the intrinsic host galaxy extinction the discrepancy would increase. Therefore the only constraint we can put on the intrinsic host galaxy absorption is that it is negligible, if we assume that the model is correct. Albeit more sophisticated models could take into account this further correction, this was beyond the goals of this work.

\section{Discussion}
\label{sec:discussion}
Once the free parameters $\nu_{\rm sa}$, $\nu_{\rm m}$, $\nu_{\rm c}$, $F_{\rm m}$ and $p$ are constrained, we can exploit the relations provided by \citet{Granot2002} to derive the global and microphysical parameters of the jet: the isotropic kinetic energy $E$, the density of the medium which surrounds the progenitor $n$, the fraction of internal energy retained by the magnetic field $\epsilon_{B}$ and the fraction of internal energy retained by the electrons $\epsilon_{e}$. From the conservation of energy we know that $\epsilon_{e} \le 1$, $\epsilon_{B} \le 1$ and $\epsilon_{e} + \epsilon_{B} \le 1$. A further constraint is given by the VHE emission: if we consider the sub-TeV emission to be due to the SSC from the relativistic electrons, then $\epsilon_{e} \ge \epsilon_{B}$ \citep{Sari2001, Zhang2001}. If we try to solve the equations from \cite{Granot2002}, the inferred parameters violate the conservation of energy, i.e. $\epsilon_{e} + \epsilon_{B} \ge 1$; however, these values are determined under the implicit assumption that all the electrons that are swept up by the forward shocks are accelerated, while this is expected to be true only for a fraction $f$ of them. As shown by \cite{Eichler2005}, if $m_{e}/m_{p} \le f \le 1$ the observed emission does not change when scaling the parameters as follows: $E \rightarrow E/f$, $\epsilon_{e} \rightarrow \epsilon_{e}f$, $\epsilon_{B} \rightarrow \epsilon_{B}f$, $n \rightarrow n/f$ \citep{vanderHorst2014}. In order to find the solutions, we make $E$ and $\epsilon_{e}$ vary within physically reasonable ranges, i.e. $10^{50}$\,erg $\le E \le$ $10^{55}$\,erg and $10^{-4} \le \epsilon_{e} \le 1$, and we subsequently calculate $\epsilon_{B}$ and $n$ using the inferred break frequencies, $F_{\rm m}$ and $p$. Finally, we apply the constraints given by the conservation of energy and the sub-TeV emission. The final solutions are listed in the second column of Table \ref{tab:quantities1}.

Furthermore, since we expect $\epsilon_{e}$ to be of the order of 0.1 from numerical simulations (\citealt{Sironi2013}, and references therein), we provide the full set of inferred values for the $0.05 \le \epsilon_{e} \le 0.15$ case in the second column of Table \ref{tab:quantities2}. We find that the isotropic kinetic energy goes from $3\times10^{50}$ to $10^{55}$\,erg. If we consider the isotropic-equivalent energy derived by \citet{Minaev2020} from the prompt emission, we can roughly estimate the efficiency of the prompt emission as $\eta = E_{iso}/(E+E_{iso})$. We estimate that $\eta\simeq$10$^{-3}$--27\%.
\begin{table}[t]
\centering
\begin{tabular}{cccc}
\toprule
Parameter    &Value &\multicolumn{2}{c}{Median}\\
             &      &ISM Sample  &RS Sample\\
\midrule
$E_{52}$/erg   &0.03 -- 10$^3$   &12   &20\\
$\epsilon_{e}$  &$10^{-4}$ -- 0.99   &0.32   &0.104\\
$\epsilon_{B}$  &8$\times10^{-7}$ -- 0.05   &2.7$\times10^{-2}$   &1.4$\times10^{-4}$\\
$n$/cm$^{-3}$   &0.4 -- 2$\times10^{4}$   &1.5   &2.15\\
$f$   &0.01 -- 1.00   &   &\\
\bottomrule
\end{tabular}
\caption[]{Global and microphysical parameters for GRB~201015A in the ISM scenario. The parameter name and the inferred value are listed in the first and second column respectively. The median of the sample by \citet{Aksulu2021} for those bursts that can be reproduced with an ISM profile is reported in the third column (ISM Sample), while the median for the sample of bursts with a claimed reverse shock component is reported in the fourth column (RS Sample).}
\label{tab:quantities1}
\end{table}
\begin{table}[t]
\centering
\begin{tabular}{cccc}
\toprule
Parameter   &Value&\multicolumn{2}{c}{Median}\\
             &      &ISM Sample  &RS Sample\\
\midrule
$E_{52}$/erg   &0.03 -- 14   &12   &20\\
$\epsilon_{e}$    &0.05 -- 0.15   &0.32   &0.104\\
$\epsilon_{B}$  &1.5$\times10^{-6}$ -- 0.05   &2.7$\times10^{-2}$   &1.4$\times10^{-4}$\\
$n$/cm$^{-3}$   &0.4 -- $10^{4}$   &1.5   &2.15\\
$f$   &0.02 -- 1.00   &   &\\
\bottomrule
\end{tabular}
\caption[]{Global and microphysical parameters for GRB~201015A in the ISM scenario if $0.05\le\epsilon_{e}\le0.15$. The parameter name and the inferred value are listed in the first and second column respectively. The median of the sample by \citet{Aksulu2021} for those bursts that can be reproduced with an ISM profile is reported in the third column (ISM Sample), while the median for the sample of bursts with a claimed reverse shock component is reported in the fourth column (RS Sample).}
\label{tab:quantities2}
\end{table}

To discuss these values in a broader context we consider a recent work by \citet{Aksulu2021}, who examined 26 GRBs with well-sampled broadband data sets. The authors found that $\epsilon_{B}$ ranges from $\approx$2.6$\times10^{-6}$ (GRB~030329) to $\approx$0.91 (GRB~130907A) for those GRBs that can be described with an ISM profile (hereafter ISM Sample), and 3 out of 13 GRBs have $\epsilon_{B} \ge$ 0.5; concerning $\epsilon_{e}$, they found a range between $\approx$0.14 (GRB~090328) and $\approx$0.89 (GRB~010222); finally, $n$ goes from $\approx$5$\times 10^{-3}$ (GRB~010222) to $\approx390$\,cm$^{-3}$ (GRB~030329).

We then consider long GRBs with a claimed RS detection (in X-rays and/or optical and/or radio) whose multi-wavelength light curves can be aptly described with an ISM profile (hereafter RS Sample): GRB~990123, 021004, 021211, 060908, 061126, 080319B, 090102 and 090424 \citep{Japelj2014}; GRB~130427A \citep{Perley2014}; GRB~160509A \citep{Laskar2016}; GRB~160625B \citep{Alexander2017}; GRB~161219B \citep{Laskar2018}; GRB~180720B \citep{Wang2019}; GRB~190829A \citep{Rhodes2020a}. 
The circumburst density for the GRBs of RS Sample goes from $\approx$5$\times10^{-5}$\,cm$^{-3}$ for GRB~160625B up to $\approx$360 cm$^{-3}$ for GRB~090201, while $\epsilon_{e}$ ranges from $\approx$4$\times 10^{-4}$ for GRB~090102 to $\approx$0.93 for 161219B, $\epsilon_{B}$ goes from $\approx$2$\times 10^{-5}$ for GRB~090102 to $\approx$0.11 for GRB~160509A.  The values we infer for GRB~201015A are therefore consistent with those found in both ISM and RS samples, even though the surrounding density is generally higher.

Finally, we consider three GRBs that have been detected at VHE: GRB~180720B, GRB~190114C and GRB~190829A. 
For these bursts $\epsilon_{e}$ goes from 0.02 (GRB~190114C; \citealt{Misra2021}) to 0.1 (GRB~180720B; \citealt{Wang2019}); $\epsilon_{B}$ goes from 4.7$\times10^{-5}$ (GRB~190114C; \citealt{Misra2021}) to $10^{-4}$ (GRB~180720B; \citealt{Wang2019}); the surrounding medium density $n$ goes from 0.1 (GRB~180720B; \citealt{Wang2019}) to 23 (GRB~190114C; \citealt{Misra2021}). These values are consistent with those we derive for GRB~201015A in this work.

From the maximum flux density $F_{\rm m}$ at 8.5\,GHz we calculate the luminosity $L$ of the afterglow with $L = F_{\rm m}4\pi d_{l}^{2}(1+z)^{\beta-\alpha-1}$ \citep{Chandra2012}, where $d_{l}$ is the luminosity distance in cm, $F_{\rm m}$ is expressed in erg\,s$^{-1}$\,cm$^{-2}$\,Hz$^{-1}$, $z$ is the redshift and $\alpha = \beta = 0$, since the peak in the light curve is also a peak in the spectrum. We find that $L\simeq3.5\times10^{30}$\,erg\,s$^{-1}$\,Hz$^{-1}$ at 1.9 days, which is slightly below the average value for radio detected GRB afterglows \citep{Chandra2012}. Finally, the maximum luminosity $L\simeq5.4\times10^{30}$\,erg\,s$^{-1}$\,Hz$^{-1}$ at 15.7\,GHz at 0.8 days is consistent with the radio luminosity previously found for the other GRBs detected at VHE \citep{Rhodes2020a}.

We stress that the allowed ranges inferred for the microphysical and global parameters of GRB~201015A are too large to pinpoint any possible deviation of this burst from the samples we used, and hence to derive important information on the production of VHE photons in GRBs. Moreover, a population study is still hindered by the paucity of GRBs detected at VHE and their proximity ($z < 1.1$), which could lead to a strong bias. A larger and more complete sample is therefore needed.

On the other hand, the fact that we cannot flag any possible deviation from the mentioned samples could be consistent with the VHE GRBs being drawn from the same parent population as the other radio-detected long GRBs \citep{Rhodes2020a}.

\subsection{Additional emission components}
\label{subsec:add_comp}
It is worth noticing that a possible refined model could be obtained by including the RS component, whose prescription could explain the bump and the observed excess in the optical emission before 0.01 days. As a matter of fact, all the GRBs with a confirmed VHE emission have been successfully modelled once a RS component was included: GRB~180720B \citep{Fraija2019, Wang2019}, GRB~190114C \citep{Laskar2019b}, GRB~190829A \citep{Rhodes2020a}.

Concerning the SN emission, if we take the emission of SN1998bw in the $r$-band \citep{Galama1998}, de-absorb the flux density using $A_{V} = 0.2$ \citep{Galama1998} and move the SN to $z = 0.426$ and 7 days earlier, we find that its light curve is consistent with that observed for GRB~201015A after 3 days from the burst (see Figure \ref{fig:MWL_LC_ISM}, dashed line). This further strengthen the SN origin of the bump observed around 10 days post-burst.

Finally, we suggest that a transition between the wind-like and the ISM profile at around 0.1 -- 0.2 days could possibly explain the change in slope observed in the X-ray light curve after $\sim$0.2 days (see, e.g., \citealt{Kamble2007, Veres2015}). As a matter of fact, the optical slope between 0.03 and 0.2 days follows a power law $F\propto t^{-1.1\pm0.2}$, which is consistent with the prediction from a model with a wind-like profile, namely $F\propto t^{-1.3}$, if the optical lies between $\nu_{m}$ and $\nu_{c}$. The prediction for the fireball model with a homogeneous circum-burst medium is $F\propto t^{-0.8}$, which is still consistent but shallower.

\subsection{High Resolution Observations}
\label{subsec:displacement}
To measure the expansion or the proper motion of the outflow, a high signal-to-noise ratio is required, as it allows both a follow-up of the afterglow up to later times and a smaller uncertainty on the position of the detected source \citep{Taylor1999}. While we achieved $\sim$mas angular resolution with EVN, we could not pinpoint any displacement of the centroid (off-axis GRB, \citealt{Mooley2018, Ghirlanda2019}) nor an expansion of the source (on-axis GRB, \citealt{Taylor2004}).

As a matter of fact, the position of the afterglow in the two detections with EVN is consistent within the uncertainties, i.e. $\Delta\alpha = 0.2$\,mas and $\Delta \delta = 0.3$\,mas. At $z=0.426$, the centroid displacement before 47 days post-burst is therefore smaller than 1.1\,pc in right ascension and 1.7\,pc in declination respectively; assuming that the burst is observed at the viewing angle $\theta$ that maximises the apparent velocity $\beta_{app} = \Gamma$, i.e. $\theta \sim \beta_{app}^{-1}$, we derive a Lorentz factor upper limit of $\Gamma_{\alpha} \le$ 40 in right ascension and $\Gamma_{\delta} \le$ 61 in declination. Considering the previous outstanding burst for which a proper motion was observed, i.e. GRB~170817A at $z=0.0093$, a displacement of the same magnitude of that of GRB~170817A would have been seen as 0.08\,mas at $z = 0.426$ after $\sim 207$ days post-burst.

On the other hand, if the GRB is seen on-axis, by taking the minor axis of the beam we constrain the size of the afterglow to be $\le$5\,pc and $\le$16\,pc at 25 and 47 days respectively. Considering the only case for which the expansion was confirmed, i.e. GRB~030329 at $z = 0.1685$, an expansion of the same magnitude of that of GRB~030329 would have been seen as 0.09\,mas at $z = 0.426$ after $\sim80$ days post-burst.

Since our best resolution with EVN is 1.8\,mas $\times$ 0.9\,mas, we would have detected such an expansion or displacement if (i) the size of the beam had not changed in later observations, (ii) the afterglow had been observable and detectable with a signal-to-noise ratio larger than 10 for about 200 days or 80 days in the case of displacement and expansion respectively and (iii) the displacement/expansion had occurred along the coordinate corresponding to the minor axis of the beam.

Conversely, considering the worst resolution reached with our VLBI observations, i.e. 3.1\,mas $\times$ 3.6\,mas, we would have pinpointed these effects if the afterglow had been detectable for about 800 days or 320 days in the case of proper motion and expansion respectively, so that the measurements to be performed would have been of the order of 0.3\,mas.

\subsection{Host Galaxy}
\label{subsec:host}
The host galaxy was firstly pinpointed by \citet{Belkin2020a} and subsequently confirmed by \citet{Rastinejad2020} and \citet{Rossi2020}, who found a magnitude $r = 22.9\pm0.2$.

With the MMT observations, we derive the position of the host: $\alpha =23^{\rm h}37^{\rm m}16.4757^{\rm s}$, $\delta = +53^{\circ}24^{\prime}54.626^{\prime\prime}$ (J2000; uncertainty = 0.235\arcsec); this is found to be 1.86$^{\prime\prime}$ far from the source observed at 1.5\,GHz, which corresponds to roughly 10\,kpc at $z = 0.426$. The uncertainty in the radio position at 1.5\,GHz is 0.03$^{\prime\prime}$, which is $\sim$170\,pc, and therefore we can state that the emission observed at 1.5\,GHz is consistent with being generated by the afterglow. Moreover, as the beam size at 1.5\,GHz is roughly 0.18$^{\prime\prime} \times$ 0.12$^{\prime\prime}$, the emitting region should be of the order of 1\,kpc $\times$ 0.7\,kpc; if the detected emission were caused by a high star forming region, we would have observed a stable emission in the optical at the same position instead of a transient event.

A safe discrimination between the galactic contamination and the proper afterglow emission at 1.5\,GHz could be achieved also with (i) a higher resolution and (ii) an improved sensitivity in late epochs, in order to get better constraints on the light curve. While the former requirement is provided by VLBI observations, the latter is reached with the Pathfinders of the Square Kilometre Array (SKA), i.e. the Meer Karoo Array Telescope (MeerKAT; see e.g. \citealt{Rhodes2020a}) and the Australian Square Kilometre Array Pathfinder (ASKAP). Moreover, a better sensitivity allows the detection of possible late time jet breaks and therefore the measurement of the jet opening angle.

\section{Conclusions}
\label{sec:conclusions}
GRB~201015A was a nearby ($z = 0.426$; \citealt{deUgartePostigo2020, Izzo2020}) long duration GRB discovered on the 15$^{\mathrm{th}}$ of October 2020 by {\em Swift}/BAT \citep{D'Elia2020a}. Its long lasting afterglow emission has been observed from $\gamma$-rays down to radio bands, claimed to be the fifth GRB ever detected at VHE energies \citep{Blanch2020a, Suda2021}.

We performed a radio follow-up with the VLA, e-MERLIN and EVN at 1.5 and 5\,GHz over twelve epochs from 1.4 to 117 days after the GRB onset. At 5\,GHz we detected a point-like source consistent with the afterglow position on 2020 October 17, 2020 November 5, 8 and 9, and 2020 December 1; conversely, on 2020 December 14, 2021 January 8 and 23, and 2021 February 9 no source was detected. At 1.5\,GHz we detected a point-like source on 2020 November 4 and 7, while on 2021 January 24 no source was detected.

We observed and detected the afterglow of GRB~201015A also in the X-rays with \emph{Chandra} (8.4 and 13.6 days post-burst) and in the optical with MMT (1.4 and 2.2 days post-burst). Finally, we collected public X-ray data from {\em Swift}/XRT and optical data from the GCN Circulars Archive. We built the multi-wavelength light curves and three spectra at 0.12, 1.41 and 23 days post-burst respectively, and we exploited the standard model provided by \cite{Granot2002} for a sharp-edged jet seen on-axis to constrain the global and microphysical parameters of the outflow: we find that the observed light curves can be reproduced with a homogeneous circumburst medium profile, and the parameters we derive for GRB~201015A are consistent with those previously found in literature for other GRBs, even though we caution that a fully reliable modelling will require a proper characterisation of the VHE detection, which is unavailable at present.

Despite the high angular resolution we achieved with the EVN observations, we could not pinpoint any change in the afterglow position. If the GRB is seen slightly off-axis, we constrain the proper motion of the outflow to be smaller than 1.1\,pc in right ascension and 1.7\,pc in declination before 47 days post-burst. Such a proper motion corresponds to a Lorentz factor upper limit of $\Gamma_{\alpha} \le$ 40 in right ascension and $\Gamma_{\delta} \le$ 61 in declination, if we assume that the GRB is seen at the viewing angle $\theta$ which maximises the apparent velocity $\beta_{app}$, i.e. $\theta \sim \beta_{app}^{-1}$. Conversely, if the GRB is seen on-axis, we find that the size of the afterglow is $\le$5\,pc and $\le$16\,pc at 25 and 47 days respectively.

We note that the bump before 0.01 days post-burst in the optical light curve could be explained by an RS component. On the other hand, we find that the \emph{Chandra} and the last {\em Swift}/XRT detections are brighter than expected from the model and from the extrapolation of the previous data points: even though further observations are needed, a late time central engine activity or a transition from a wind-like profile to a homogeneous surrounding medium at early times could possibly explain the change in the slope of the X-ray light curve.

\begin{acknowledgements}
The authors would like to thank the anonymous referee for their helpful comments.

e-MERLIN is a National Facility operated by the University of Manchester at Jodrell Bank Observatory on behalf of STFC, part of UK Research and Innovation.

The European VLBI Network is a joint facility of independent European, African, Asian, and North American radio astronomy institutes. Scientific results from data presented in this publication are derived from the following EVN project code: RM016. We thank the directors and staff of all the EVN telescopes for making this target of opportunity observation possible.

The National Radio Astronomy Observatory is a facility of the National Science Foundation operated under cooperative agreement by Associated Universities, Inc.

MMT Observatory access was supported by Northwestern University and the Center for Interdisciplinary Exploration and Research in Astrophysics (CIERA).

The scientific results reported in this article are based in part on observations made by the Chandra X-ray Observatory (PI: Gompertz; project code: 22400511). This research has made use of software provided by the Chandra X-ray Center (CXC) in the application packages CIAO and Sherpa.

BM and JMP acknowledge financial support from the State Agency for Research of the Spanish Ministry of Science and Innovation
under grant PID2019-105510GB-C31 and through the Unit of Excellence Mar\'ia de Maeztu 2020-2023 award to the Institute of Cosmos Sciences (CEX2019-000918-M).

AJL has received funding from the European Research Council (ERC) under the European Union’s Seventh Framework Programme (FP7-2007-2013) (Grant agreement No. 725246).

BPG acknowledges funding from the European Research Council (ERC) under the European Union’s Horizon 2020 research and innovation programme (grant agreement No.~948381, PI: Nicholl).
\end{acknowledgements}

%

\begin{thebibliography}{}

\bibitem[Abdalla et al.(2019)]{Abdalla2019} Abdalla, H., Adam, R., Aharonian, F., et al.\ 2019, \nat, 575, 464

\bibitem[Ackley et al.(2020)]{Ackley2020} Ackley, K., Galloway, D.~K., Mong, Y.-L., et al.\ 2020, GRB Coordinates Network, Circular Service, No. 28639, 28639

\bibitem[Aksulu et al.(2021)]{Aksulu2021} Aksulu, M.~D., Wijers, R.~A.~M.~J., van Eerten, H.~J., et al.\ 2021, arXiv:2106.14921

\bibitem[Alexander et al.(2017)]{Alexander2017} Alexander, K.~D., Laskar, T., Berger, E., et al.\ 2017, \apj, 848, 69

\bibitem[Amati et al.(2002)]{Amati2002} Amati, L., Frontera, F., Tavani, M., et al.\ 2002, \aap, 390, 81

\bibitem[Beasley et al.(2002)]{beasley2002} Beasley A.~J., Gordon D., Peck A.~B., Petrov L., MacMillan D.~S., Fomalont E.~B., Ma C.\ 2002, ApJS, 141, 13.

\bibitem[Becker(2015)]{Becker15} Becker, A.\ 2015, Astrophysics Source Code Library. ascl:1504.00

\bibitem[Belkin et al.(2020a)]{Belkin2020a} Belkin, S., Pankov, N., Pozanenko, A., et al.\ 2020, GRB Coordinates Network, Circular Service, No. 28656, 28656

\bibitem[Belkin et al.(2020b)]{Belkin2020b} Belkin, S., Kim, V., Pozanenko, A., et al.\ 2020, GRB Coordinates Network, Circular Service, No. 28673, 28673

\bibitem[Berger et al.(2004)]{Berger2004} Berger, E., Kulkarni, S.~R., \& Frail, D.~A.\ 2004, \apj, 612, 966

\bibitem[Berger(2014)]{Berger2014} Berger, E.\ 2014, \araa, 52, 43

\bibitem[Blanch et al.(2020a)]{Blanch2020a} Blanch, O., Gaug, M., Noda, K., et al.\ 2020, GRB Coordinates Network, Circular Service, No. 28659, 28659

\bibitem[Blanch et al.(2020b)]{Blanch2020b} Blanch, O., Longo, F., Berti, A., et al.\ 2020, GRB Coordinates Network, Circular Service, No. 29075, 29075

\bibitem[Chambers et al.(2016)]{Chambers+16} Chambers, K.~C., Magnier, E.~A., Metcalfe, N., et al.\ 2016, arXiv:1612.0556

\bibitem[Chandra \& Frail(2012)]{Chandra2012} Chandra, P. \& Frail, D.~A.\ 2012, \apj, 746, 156.

\bibitem[Condon(1992)]{Condon1992} Condon, J.~J.\ 1992, \araa, 30, 575.

\bibitem[D'Ai et al.(2020)]{DAi2020} D'Ai, A., Gropp, J.~D., Kennea, J.~A., et al.\ 2020, GRB Coordinates Network, Circular Service, No. 28660, 28660

\bibitem[D'Elia et al.(2020)]{D'Elia2020a} D'Elia, V., Ambrosi, E., Barthelmy, S.~D., et al.\ 2020, GRB Coordinates Network, Circular Service, No. 28632, 28632

\bibitem[D'Elia \& Swift Team(2020)]{D'Elia2020b} D'Elia, V. \& Swift Team\ 2020, GRB Coordinates Network, Circular Service, No. 28857, 28857

\bibitem[de Ugarte Postigo et al.(2020)]{deUgartePostigo2020} de Ugarte Postigo, A., Kann, D.~A., Blazek, M., et al.\ 2020, GRB Coordinates Network, Circular Service, No. 28649, 28649

\bibitem[Drenkhahn \& Spruit(2002)]{Drenkhahn2002} Drenkhahn, G. \& Spruit, H.~C.\ 2002, \aap, 391, 1141

\bibitem[Eichler \& Waxman(2005)]{Eichler2005} Eichler, D. \& Waxman, E.\ 2005, \apj, 627, 861

\bibitem[Evans et al.(2007)]{Evans2007} Evans, P.~A., Beardmore, A.~P., Page, K.~L., et al.\ 2007, \aap, 469, 379

\bibitem[Evans et al.(2009)]{Evans2009} Evans, P.~A., Beardmore, A.~P., Page, K.~L., et al.\ 2009, \mnras, 397, 1177

\bibitem[Fletcher et al.(2020)]{Fletcher2020} Fletcher, C., Veres, P., \& Fermi-GBM Team\ 2020, GRB Coordinates Network, Circular Service, No. 28663, 28663

\bibitem[Fong et al.(2020)]{Fong2020} Fong, W., Schroeder, G., Rastinejad, J., et al.\ 2020, GRB Coordinates Network, Circular Service, No. 28688, 28688

\bibitem[Fraija et al.(2019)]{Fraija2019} Fraija, N., Dichiara, S., Pedreira, A.~C.~C. do E.~S., et al.\ 2019, \apj, 885, 29

\bibitem[Fraija et al.(2021)]{Fraija2021} Fraija, N., Veres, P., Beniamini, P., et al.\ 2021, \apj, 918, 12

\bibitem[Frail et al.(1997)]{Frail1997} Frail, D.~A., Kulkarni, S.~R., Nicastro, L., et al.\ 1997, \nat, 389, 261

\bibitem[Frail et al.(2004)]{Frail2004} Frail, D.~A., Metzger, B.~D., Berger, E., et al.\ 2004, \apj, 600, 828

\bibitem[Frail et al.(2005)]{Frail2005} Frail, D.~A., Soderberg, A.~M., Kulkarni, S.~R., et al.\ 2005, \apj, 619, 994

\bibitem[Galama et al.(1998)]{Galama1998} Galama, T.~J., Vreeswijk, P.~M., van Paradijs, J., et al.\ 1998, \nat, 395, 670.

\bibitem[Geng et al.(2018)]{Geng2018} Geng, J.-J., Dai, Z.-G., Huang, Y.-F., et al.\ 2018, \apjl, 856, L33

\bibitem[Gehrels et al.(2009)]{Gehrels2009} Gehrels, N., Ramirez-Ruiz, E., \& Fox, D.~B.\ 2009, \araa, 47, 567.

\bibitem[Ghirlanda et al.(2019)]{Ghirlanda2019} Ghirlanda, G., Salafia, O.~S., Paragi, Z., et al.\ 2019, Science, 363, 968.

\bibitem[Giarratana et al.(2020)]{Giarratana2020} Giarratana, S., Giroletti, M., Marcote, B., et al.\ 2020, GRB Coordinates Network, Circular Service, No. 28939, 28939

\bibitem[Gompertz et al.(2020)]{Gompertz2020} Gompertz, B., Levan, A., Tanvir, N., et al.\ 2020, GRB Coordinates Network, Circular Service, No. 28822, 28822

\bibitem[Gordon et al.(2016)]{gordon2016} Gordon D., Jacobs C., Beasley A., Peck A., Gaume R., Charlot P., Fey A., et al.\ 2016, AJ, 151, 154.

\bibitem[Granot \& Sari(2002)]{Granot2002} Granot, J. \& Sari, R.\ 2002, \apj, 568, 820

\bibitem[Greisen(2003)]{Greisen2003} Greisen, E.~W.\ 2003, Information Handling in Astronomy - Historical Vistas, 109

\bibitem[Grossan et al.(2020)]{Grossan2020} Grossan, B., Maksut, Z., Kim, A., et al.\ 2020, GRB Coordinates Network, Circular Service, No. 28674, 28674

\bibitem[H.~E.~S.~S. Collaboration et al.(2021)]{HESS2021} H.~E.~S.~S. Collaboration, Abdalla, H., Aharonian, F., et al.\ 2021, Science, 372, 1081

\bibitem[Hinshaw et al.(2013)]{Hinshaw2013} Hinshaw, G., Larson, D., Komatsu, E., et al.\ 2013, \apjs, 208, 19

\bibitem[Hu et al.(2020)]{Hu2020} Hu, Y.-D., Fernandez-Garcia, E., Castro-Tirado, A.~J., et al.\ 2020, GRB Coordinates Network, Circular Service, No. 28645, 28645

\bibitem[Izzo et al.(2020)]{Izzo2020} Izzo, L., Malesani, D.~B., Zhu, Z.~P., et al.\ 2020, GRB Coordinates Network, Circular Service, No. 28661, 28661

\bibitem[Japelj et al.(2014)]{Japelj2014} Japelj, J., Kopa{\v{c}}, D., Kobayashi, S., et al.\ 2014, \apj, 785, 84

\bibitem[Jelinek et al.(2020)]{Jelinek2020} Jelinek, M., Strobl, J., Karpov, S., et al.\ 2020, GRB Coordinates Network, Circular Service, No. 28664, 28664

\bibitem[Jin et al.(2013)]{Jin2013} Jin, Z.-P., Covino, S., Della Valle, M., et al.\ 2013, \apj, 774, 114

\bibitem[Kalberla et al.(2005)]{Kalberla2005} Kalberla, P.~M.~W., Burton, W.~B., Hartmann, D., et al.\ 2005, \aap, 440, 775

\bibitem[Kamble et al.(2007)]{Kamble2007} Kamble, A., Resmi, L., \& Misra, K.\ 2007, \apjl, 664, L5

\bibitem[Kann et al.(2006)]{Kann2006} Kann, D.~A., Klose, S., \& Zeh, A.\ 2006, \apj, 641, 993

\bibitem[Keimpema et al.(2015)]{Keimpema2015} Keimpema, A., Kettenis, M.~M., Pogrebenko, S.~V., et al.\ 2015, Experimental Astronomy, 39, 259

\bibitem[Kennea et al.(2020)]{Kennea2020} Kennea, J.~A., Tagliaferri, G., Campana, S., et al.\ 2020, GRB Coordinates Network, Circular Service, No. 28635, 28635

\bibitem[Kouveliotou et al.(1993)]{Kouveliotou1993} Kouveliotou, C., Meegan, C.~A., Fishman, G.~J., et al.\ 1993, \apjl, 413, L101

\bibitem[Kouveliotou et al.(2012)]{Kouveliotou2012} Kouveliotou, C., Wijers, R.~A.~M.~J., \& Woosley, S.\ 2012, Gamma-ray Bursts, by Chryssa Kouveliotou , Ralph A. M. J. Wijers , Stan Woosley, Cambridge, UK: Cambridge University Press, 2012
\bibitem[Kumar \& Zhang(2015)]{Kumar2015} Kumar, P. \& Zhang, B.\ 2015, \physrep, 561.

\bibitem[Kumar et al.(2020a)]{Kumar2020a} Kumar, H., Sahu, D.~K., Gupta, R., et al.\ 2020, GRB Coordinates Network, Circular Service, No. 28680, 28680

\bibitem[Kumar et al.(2020b)]{Kumar2020b} Kumar, H., Stanzin, U., Bhalerao, V., et al.\ 2020, GRB Coordinates Network, Circular Service, No. 28681, 28681

\bibitem[Laskar et al.(2016)]{Laskar2016} Laskar, T., Alexander, K.~D., Berger, E., et al.\ 2016, \apj, 833, 88

\bibitem[Laskar et al.(2018)]{Laskar2018} Laskar, T., Alexander, K.~D., Berger, E., et al.\ 2018, \apj, 862, 94

\bibitem[Laskar et al.(2019b)]{Laskar2019b} Laskar, T., Alexander, K.~D., Gill, R., et al.\ 2019, \apjl, 878, L26

\bibitem[Lipunov et al.(2020a)]{Lipunov2020a} Lipunov, V., Gorbovskoy, E., Kornilov, V., et al.\ 2020, GRB Coordinates Network, Circular Service, No. 28633, 28633

\bibitem[Lipunov et al.(2020b)]{Lipunov2020b} Lipunov, V., Gorbovskoy, E., Kornilov, V., et al.\ 2020, GRB Coordinates Network, Circular Service, No. 28634, 28634

\bibitem[MAGIC Collaboration et al.(2019)]{Magic2019} MAGIC Collaboration, Acciari, V.~A., Ansoldi, S., et al.\ 2019, \nat, 575, 455.

\bibitem[Malesani et al.(2020)]{Malesani2020} Malesani, D.~B., de Ugarte Postigo, A., \& Pursimo, T.\ 2020, GRB Coordinates Network, Circular Service, No. 28637, 28637

\bibitem[Marcote et al.(2020)]{Marcote2020} Marcote, B., Rib{\'o}, M., Paredes, J.~M., et al.\ 2020, GRB Coordinates Network, Circular Service, No. 29028, 29028

\bibitem[Markwardt et al.(2020)]{Markwardt2020} Markwardt, C.~B., Barthelmy, S.~D., Cummings, J.~R., et al.\ 2020, GRB Coordinates Network, Circular Service, No. 28658, 28658

\bibitem[Marshall et al.(2020)]{Marshall2020} Marshall, F.~E., D'Elia, V., \& Swift/UVOT Team\ 2020, GRB Coordinates Network, Circular Service, No. 28662, 28662

\bibitem[Mazets et al.(1981)]{Mazets1981} Mazets, E.~P., Golenetskii, S.~V., Ilyinskii, V.~N., et al.\ 1981, \apss, 80, 119

\bibitem[McMullin et al.(2007)]{McMullin2007} McMullin, J.~P., Waters, B., Schiebel, D., et al.\ 2007, Astronomical Data Analysis Software and Systems XVI, 376, 127

\bibitem[M{\'e}sz{\'a}ros \& Rees(1993)]{Meszaros1993} M{\'e}sz{\'a}ros, P. \& Rees, M.~J.\ 1993, \apj, 405, 278

\bibitem[M{\'e}sz{\'a}ros \& Rees(1994)]{Meszaros1994} M{\'e}sz{\'a}ros, P. \& Rees, M.~J.\ 1994, \mnras, 269, L41

\bibitem[M{\'e}sz{\'a}ros et al.(1994)]{Meszaros1994b} M{\'e}sz{\'a}ros, P., Rees, M.~J., \& Papathanassiou, H.\ 1994, \apj, 432, 181

\bibitem[M{\'e}sz{\'a}ros(2002)]{Meszaros2002} M{\'e}sz{\'a}ros, P.\ 2002, \araa, 40, 137

\bibitem[Minaev \& Pozanenko(2020)]{Minaev2020} Minaev, P. \& Pozanenko, A.\ 2020, GRB Coordinates Network, Circular Service, No. 28668, 28668

\bibitem[Misra et al.(2021)]{Misra2021} Misra, K., Resmi, L., Kann, D.~A., et al.\ 2021, \mnras, 504, 5685

\bibitem[Monet et al.(2003)]{USNO-B1} Monet, D.~G., Levine, S.~E., Canzian, B., et al.\ 2003, \aj, 125, 984

\bibitem[Mooley et al.(2018)]{Mooley2018} Mooley, K.~P., Deller, A.~T., Gottlieb, O., et al.\ 2018, \nat, 561, 355

\bibitem[Moskvitin et al.(2020)]{Moskvitin2020} Moskvitin, A.~S., Aitov, V.~N., \& GRB follow-up Team\ 2020, GRB Coordinates Network, Circular Service, No. 28721, 28721

\bibitem[Norris et al.(1984)]{Norris1984} Norris, J.~P., Cline, T.~L., Desai, U.~D., et al.\ 1984, \nat, 308, 434

\bibitem[Paczy{\'n}ski(1998)]{Paczynski1998} Paczy{\'n}ski, B.\ 1998, \apjl, 494, L45

\bibitem[Perley et al.(2014)]{Perley2014} Perley, D.~A., Cenko, S.~B., Corsi, A., et al.\ 2014, \apj, 781, 37.

\bibitem[Piran(2004)]{Piran2004} Piran, T.\ 2004, Reviews of Modern Physics, 76, 1143

\bibitem[Pozanenko et al.(2020)]{Pozanenko2020} Pozanenko, A., Belkin, S., Volnova, A., et al.\ 2020, GRB Coordinates Network, Circular Service, No. 29033, 29033

\bibitem[Pradel, et al.(2006)]{pradel2006} Pradel N., Charlot P., Lestrade J.-F.\ 2006, A\&A, 452, 1099.

\bibitem[Rastinejad et al.(2020)]{Rastinejad2020} Rastinejad, J., Paterson, K., Kilpatrick, C.~D., et al.\ 2020, GRB Coordinates Network, Circular Service, No. 28676, 28676

\bibitem[Rees \& M{\'e}sz{\'a}ros(1992)]{Rees1992} Rees, M.~J. \& M{\'e}sz{\'a}ros, P.\ 1992, \mnras, 258, 41

\bibitem[Rees \& M{\'e}sz{\'a}ros(1994)]{Rees1994} Rees, M.~J. \& M{\'e}sz{\'a}ros, P.\ 1994, \apjl, 430, L93

\bibitem[Rhodes et al.(2020b)]{Rhodes2020b} Rhodes, L., Fender, R., Bray, J., et al.\ 2020, GRB Coordinates Network, Circular Service, No. 28945, 28945

\bibitem[Rhodes et al.(2020a)]{Rhodes2020a} Rhodes, L., van der Horst, A.~J., Fender, R., et al.\ 2020, \mnras, 496, 3326

\bibitem[Rossi et al.(2021)]{Rossi2020} Rossi, A., Benetti, S., Palazzi, E., et al.\ 2021, GRB Coordinates Network, Circular Service, No. 29306, 29306

\bibitem[Salafia et al.(2021)]{Salafia2021} Salafia, O.~S., Ravasio, M.~E., Yang, J., et al.\ 2021, arXiv:2106.07169

\bibitem[Sari \& Esin(2001)]{Sari2001} Sari, R. \& Esin, A.~A.\ 2001, \apj, 548, 787

\bibitem[Schlafly \& Finkbeiner(2011)]{Schlafly2011} Schlafly, E.~F. \& Finkbeiner, D.~P.\ 2011, \apj, 737, 103

\bibitem[Selsing et al.(2019)]{Selsing2019} Selsing, J., Fynbo, J.~P.~U., Heintz, K.~E., et al.\ 2019, GRB Coordinates Network, Circular Service, No. 23695, 23695

\bibitem[Shepherd et al.(1994)]{Shepherd1994} Shepherd, M.~C., Pearson, T.~J., \& Taylor, G.~B.\ 1994, \baas

\bibitem[Sironi et al.(2013)]{Sironi2013} Sironi, L., Spitkovsky, A., \& Arons, J.\ 2013, \apj, 771, 54

\bibitem[Suda et al.(2021)]{Suda2021} Suda, Y., Artero, M., Asano, K. et al.\ 2021, Proceedings of 37th International Cosmic Ray Conference {\textemdash} PoS(ICRC2021), 395, 797

\bibitem[Taylor et al.(2004)]{Taylor2004} Taylor, G.~B., Frail, D.~A., Berger, E., et al.\ 2004, \apjl, 609, L1

\bibitem[Taylor et al.(1999)]{Taylor1999} Taylor, G.~B., Carilli, C.~L., \& Perley, R.~A.\ 1999, Synthesis Imaging in Radio Astronomy II, 180

\bibitem[Tody(1993)]{Tody93} Tody, D.\ 1993, Astronomical Data Analysis Software and Systems II, 52, 173

\bibitem[Usov(1992)]{Usov1992} Usov, V.~V.\ 1992, \nat, 357, 472

\bibitem[Valeev et al.(2019)]{Valeev2019} Valeev, A.~F., Castro-Tirado, A.~J., Hu, Y.-D., et al.\ 2019, GRB Coordinates Network, Circular Service, No. 25565, 25565

\bibitem[van der Horst et al.(2014)]{vanderHorst2014} van der Horst, A.~J., Paragi, Z., de Bruyn, A.~G., et al.\ 2014, \mnras, 444, 3151

\bibitem[Veres et al.(2015)]{Veres2015} Veres, P., Corsi, A., Frail, D.~A., et al.\ 2015, \apj, 810, 31

\bibitem[Vielfaure et al.(2020)]{Vielfaure2020} Vielfaure, J.-B., Izzo, L., Xu, D., et al.\ 2020, GRB Coordinates Network, Circular Service, No. 29077, 29077

\bibitem[Vreeswijk et al.(2018)]{Vreeswijk2018} Vreeswijk, P.~M., Kann, D.~A., Heintz, K.~E., et al.\ 2018, GRB Coordinates Network, Circular Service, No. 22996, 22996

\bibitem[Wang et al.(2019)]{Wang2019} Wang, X.-Y., Liu, R.-Y., Zhang, H.-M., et al.\ 2019, \apj, 884, 117
\bibitem[Woosley(1993)]{Woosley1993} Woosley, S.~E.\ 1993, \apj, 405, 273

\bibitem[Willingale et al.(2013)]{Willingale2013} Willingale, R., Starling, R.~L.~C., Beardmore, A.~P., et al.\ 2013, \mnras, 431, 394

\bibitem[Wilms et al.(2000)]{Wilms2000} Wilms, J., Allen, A., \& McCray, R.\ 2000, \apj, 542, 914

\bibitem[Woosley \& Heger(2006)]{Woosley2006} Woosley, S.~E. \& Heger, A.\ 2006, \apj, 637, 914

\bibitem[Zhang \& M{\'e}sz{\'a}ros(2001)]{Zhang2001} Zhang, B. \& M{\'e}sz{\'a}ros, P.\ 2001, \apj, 559, 110

\bibitem[Zhang \& M{\'e}sz{\'a}ros(2004)]{Zhang2004} Zhang, B. \& M{\'e}sz{\'a}ros, P.\ 2004, International Journal of Modern Physics A, 19, 2385

\bibitem[Zhang et al.(2021)]{Zhang2021} Zhang, L.-L., Ren, J., Huang, X.-L., et al.\ 2021, \apj, 917, 95

\bibitem[Zhang \& MacFadyen(2009)]{Zhang2009} Zhang, W. \& MacFadyen, A.\ 2009, \apj, 698, 1261

\bibitem[Zhu et al.(2020a)]{Zhu2020a} Zhu, Z.~P., Liu, X., Fu, S.~Y., et al.\ 2020, GRB Coordinates Network, Circular Service, No. 28653, 28653

\bibitem[Zhu et al.(2020b)]{Zhu2020b} Zhu, Z.~P., Liu, X., Fu, S.~Y., et al.\ 2020, GRB Coordinates Network, Circular Service, No. 28677, 28677
\end{thebibliography}
%

\end{document}